\documentclass{aa}  

\usepackage{graphicx}
\usepackage{txfonts}
\usepackage[colorlinks,citecolor=blue]{hyperref}
\hypersetup{
    colorlinks = true,
    linkcolor = blue,
    anchorcolor = blue,
    citecolor = blue,
    filecolor = blue,
    urlcolor = blue
    }

\usepackage{booktabs} 
\usepackage{mathrsfs}
\usepackage{tikz}
\usepackage{booktabs}
\usetikzlibrary{calc}
\usepackage{mathtools}
\usepackage{siunitx}
\usepackage{svg}

\setcitestyle{notesep={},round,aysep={},yysep={;}}

\begin{document} 

   \title{The blue supergiant Sher 25 revisited in the Gaia era}

   \author{D. We{\ss}mayer\inst{1}
          \and
          N. Przybilla\inst{1}
          \and
          A. Ebenbichler\inst{1}
          \and
          P. Aschenbrenner\inst{1}
          \and
          K. Butler\inst{2}
          }

   \institute{Universit\"at Innsbruck, Institut f\"ur Astro- und Teilchenphysik, Technikerstr. 25/8, 6020 Innsbruck, Austria\\
              \email{david.wessmayer@uibk.ac.at ; norbert.przybilla@uibk.ac.at}
         \and
             Ludwig-Maximilians-Universit\"at M\"unchen, Universit\"atssternwarte, Scheinerstr. 1, 81679 M\"unchen, Germany
             }

   \date{Received ; accepted }

 
  \abstract
  {}
   {
   The evolutionary status of the blue supergiant Sher~25 and its membership to the massive cluster NGC~3603 are investigated.
   }
   {A hybrid non-LTE (local thermodynamic equilibrium) spectrum synthesis approach is employed to analyse a high-resolution optical spectrum of Sher~25 and five similar early B-type comparison stars in order to derive atmospheric parameters and elemental abundances. Fundamental stellar parameters are determined by considering stellar evolution tracks, Gaia Data Release 3 (DR3) data and complementary distance information. Interstellar reddening and the reddening law along the sight line towards Sher~25 are constrained employing UV photometry for the first time in addition to optical and infrared data. 
   The distance to NGC~3603 is reevaluated based on Gaia DR3 data of the innermost cluster O-stars. 
   }
   {The spectroscopic distance derived from the quantitative analysis implies that Sher~25 lies in the foreground of NGC~3603, which is found to have a distance of $d_\mathrm{NGC 3603}$\,=\,6250$\pm$150\,pc. A cluster membership is also excluded as the hourglass nebula is unaffected by the vigorous stellar winds of the cluster stars and from the different excitation signatures of the hourglass nebula and the nebula around NGC~3603. Sher~25 turns out to have a luminosity of $\log L/L_\odot$\,=\,5.48$\pm$0.14, equivalent to that of a $\sim$27\,$M_\odot$ supergiant in a single-star scenario, which is about half of the mass assumed so far, bringing it much closer in its characteristics to Sk$-$69\degr202, the progenitor of SN~1987A. Sher~25 is significantly older than NGC~3603. Further arguments for a binary (merger) evolutionary scenario of Sher~25 are discussed.
   } 
   {}


   \maketitle
%

\section{Introduction}
\begin{figure}
\centering
\includegraphics[width=0.95\hsize]{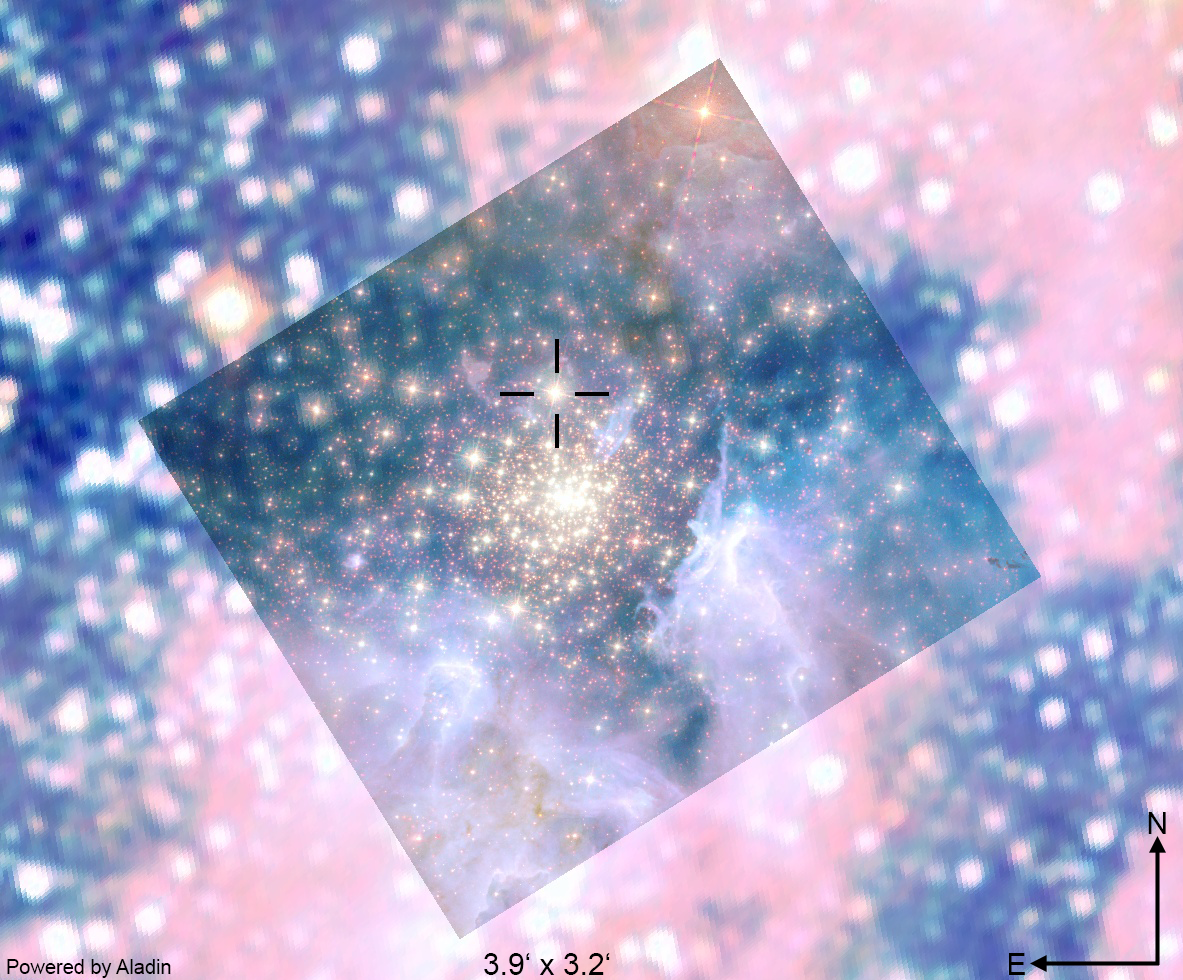}
\caption{Colour composite of the field of NGC~3603, with Sher~25 marked by the crosshairs. The centre shows 
a semi-transparent Hubble Space Telescope Wide Field Camera 3 (HST/WFC3) image (proposal ID: 11360, PI:
Robert O’Connel) with the following colour coding: blue (F656N filter), green (F673N), yellow (F128N) and red 
(F164N). The background shows a 2MASS colour image composed of $J$ (1.235\,$\mu$m, blue), $H$ (1.662\,$\mu$m, green) 
and $K_\mathrm{s}$ exposures (2.159\,$\mu$m, red).}
\label{fig:ngc3603_field}
\end{figure}

The star \object{Sher 25} \citep{Sher65} is an intriguing blue supergiant (BSG) close to the massive 
Galactic cluster \object{NGC 3603}. It is located at about 20\arcsec~distance north of the cluster centre, see 
Fig.~\ref{fig:ngc3603_field} for a composite image of the field, combining optical and near infrared observations 
with the Hubble Space Telescope (HST) with near-infrared observations obtained within the Two Micron All Sky Survey 
(2MASS). The sight line is complex and crowded, aligned over a long stretch of the Carina-Sagittarius spiral arm and 
characterised by varying reddening towards individual stars and variable nebular emission throughout the field 
\citep{Pangetal11}. 
Sher~25 shows a surrounding hourglass-shaped nebula \citep{Brandner97a,Brandner97b}, similar to the triple ring 
nebula expelled by the precursor of \object{SN 1987A} \citep{Wampleretal90,Burrowsetal95}. This nebula, in 
combination with the spectral similarity of Sher~25 \citep[classified as B1~Iab,][]{Melenaetal08} to the SN~1987A 
progenitor star Sk$-$69\degr202  \citep[classified as B0.7-3\,I,][]{Walbornetal89}, raised the prospect of studying a near-
twin to a core-collapse supernova progenitor with modern observational data and analysis tools. 

\citet{Smarttetal02} reported the first model atmosphere analysis of Sher~25. They employed the {\sc Tlusty} code
\citep{Hubeny88}, where deviations from the assumption of local thermodynamic equilibrium (LTE) -- so-called non-LTE effects -- 
were allowed for in the atmospheric structure computation, but metal-blanketing effects were neglected. 
On the basis of the model atmospheres, non-LTE line-formation calculations with 
earlier versions of the codes as used here (see Sect.~\ref{section:models}) were conducted. 
Sher~25 was found to be highly luminous (adopting a distance to NGC~3603 of 6.3\,kpc), with $\log\,L/L_\odot$\,=\,5.9$\pm$0.2,
implying a zero-age main sequence (ZAMS) mass of around 60\,$M_\odot$. The surface CNO abundances were found to show some mixing with 
nuclear-processed material but to be incompatible with a previous red supergiant (RSG) phase of Sher~25 where highly-efficient 
convective dredge-up would have occurred, such that the nebula was likely ejected during the BSG phase. The overall picture 
put Sher~25 closer in nature to the high-mass Luminous Blue Variables (LBVs) than to the precursor of SN~1987A. Some LBVs produce 
ring nebulae as found around the candidate LBVs \object{HD 168625} \citep{Smith07} and \object{[SBW2007] 1} \citep{Smithetal07}, see 
also \citet{Weis11}.

The reanalysis of Sher~25 with metal line-blanketed hydrostatic ({\sc Tlusty}) and unified (photoshere+wind) non-LTE model atmospheres 
({\sc Fastwind}, \citealt{Pulsetal05}; {\sc Cmfgen}, \citealt{HiMi98}) by \citet{Hendryetal08} 
confirmed the earlier findings. Using a refined distance to NGC~3603 of 7.6\,kpc and a consistent line-of-sight extinction \citep{Melenaetal08}
a luminosity of $\log\,L/L_\odot$\,=\,5.78 was adopted, indicating a ZAMS mass of 50$\pm$10\,$M_\odot$. In addition, the
quantitative analysis of the surrounding nebula by \citet{Hendryetal08} confirmed its highly nitrogen-rich composition, while the 
oxygen abundance resembled that of the NGC~3603 background nebula. Radial velocity variations reported by \citet{Hendryetal08}
were later suggested to be due to pulsations of Sher~25 and not due to binarity \citep{Tayloretal14}.

With regard to the origin of hourglass or ring nebulae, a binary merger scenario is likely for the precursor of SN~1987A
\citep{Podsiadlowski92,MoPo07}, with the nebula shaped by the interaction of the common-envelope ejecta with the fast stellar wind of
the BSG resulting from the merger. Alternatively, a single-star scenario for ring nebula production may also be at work 
\citep{Chitaetal08}, where the fast anisotropic stellar wind of a BSG on a blue loop catches up with the slow spherical wind expelled 
during a previous RSG phase. Both scenarios require that the star which ejected the nebula had reached the RSG stage earlier in its 
life, whereas high-mass LBVs eject nebulae as BSGs \citep{Lamersetal01}. 

The parent cluster of Sher~25, NGC~3603, also deserves a few remarks. Located in the Carina-Sagittarius spiral arm, it 
is the densest concentration of massive stars known in the Milky Way. It contains several dozen early O-type stars and four WN6h stars
(including two in one of the most massive binaries of the Milky Way) that power the surrounding giant \ion{H}{ii} region. 
The starburst cluster has an age of 1$\pm$1\,Myr \citep{SuBe04,Melenaetal08} and shows ongoing star formation in the surrounding nebular knots. There are also 
indications for subsequent star formation, as Sher~25 and a second BSG, \object{Sher 23}, are about 3\,Myr older than the stars at 
the cluster centre \citep{Melenaetal08}.

However, when taking a closer look, one finds contradictions in the overall picture. If Sher~25 is indeed a member of NGC~3603, 
why is the hourglass nebula unaffected by the stellar winds from the cluster members that have otherwise cleared the sight line
towards the cluster and have shaped cold molecular pillars at even larger lateral distances from the cluster centre, see
Fig.~\ref{fig:ngc3603_field}? Why does the hourglass nebula show a low-excitation emission spectrum despite 
supposedly being in the
presence of one of the most intense UV environments known in the Milky Way, while the spectrum of the surrounding \ion{H}{ii} region 
shows the expected high excitation? Why is Sher~25 located close to the evolutionary track of a $\sim$30\,$M_\odot$ ZAMS star in a Kiel 
diagram (surface gravity vs. effective temperature), whereas it is close to a $\sim$50\,$M_\odot$ track
in the Hertzsprung-Russell diagram (HRD)? 

The answer may lie in the complex line-of-sight towards NGC~3603, which traverses the Carina-Sagittarius arm over a range of many kpc. 
A location of Sher~25 in the fore- or background of the cluster could solve the contradictions and, even more, it would also remove the 
necessity for subsequent star-formation having occurred in NGC~3603. We therefore employ data from the Gaia Data Release 3 
\citep[DR3,][]{Gaia2016,GaiaDR3} to re-investigate the cluster membership of Sher~25, and our recently introduced hybrid non-LTE 
spectral analysis methodology \citep[, henceforth Paper~I]{Wessmayeretal22} to readdress its evolutionary status.

The paper is organised as follows: the observational material on Sher~25 and the comparison stars is
summarised in Sect.~\ref{section:observations} and Sect.~\ref{section:models} concentrates on the models and the 
analysis methodology. The analysis results are presented in Sect.~\ref{section:results}, a summary of the individual comparison stars is given in Sect.~\ref{section:individual} and a discussion of Sher~25 in Sect.~\ref{section:discussion}. An IR excess along the sight line to the cluster NGC~3603 is discussed in Appendix~\ref{appendix:A}, a Gaia-based distance to NGC~3603 is derived in Appendix~\ref{appendix:B} and the model fit to the observed spectrum of Sher~25 is visualised in Appendix~\ref{appendix:C}.

\begin{figure*}
\centering 
    \resizebox{0.87\textwidth}{!}{\includegraphics{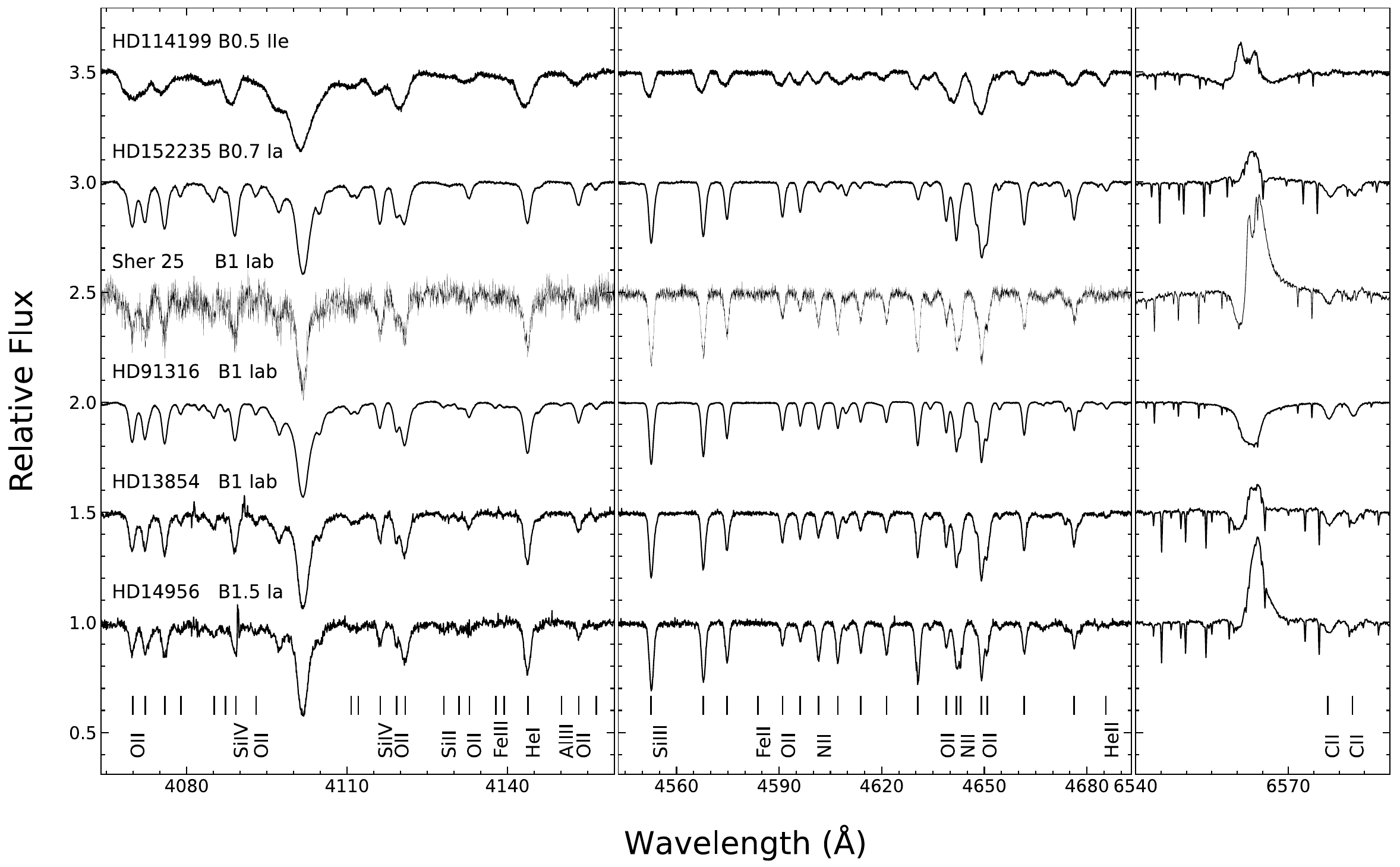}}
      \caption{Sample spectra ordered with respect to spectral type. 
      The panels show spectral windows with prominent features in early B-type supergiants.} 
         \label{fig:spect_lum_showcase_new}
\end{figure*}

\begin{table*}
\caption{B-type supergiant sample.}
\label{tab:spectra_info}
\centering  
{\small
\setlength{\tabcolsep}{0.85mm}
\begin{tabular}{llllrrlcccccc}
\hline\hline
ID   & Object             & Sp. T.                    & OB Assoc./                &$V$\tablefootmark{a}&$B-V$\tablefootmark{a}& Instrument & Date of Obs. & $T_{\mathrm{exp}}$ & $S/N$ & & \multicolumn{2}{c}{IUE}\\ \cline{12-13}
\#   &                    &                           & Cluster                   &              mag   &                mag   &            & YYYY-MM-DD   &                  s &       & &  SW    & LW            \\ \hline
1    & \object{Sher 25}   & B1\,Iab\tablefootmark{b}  & NGC 3603\tablefootmark{c} & 12.275$\pm$0.005   &    1.369$\pm$0.010   & FEROS      & 2009-03-22   & 3$\times$2400      & 160   & &  ...   & ... \\
2    & \object{HD 13854}  & B1\,Iab\tablefootmark{d}  & Per OB1                   &  6.473$\pm$0.017   &    0.281$\pm$0.003   & FOCES      & 2005-09-25   & 1800               & 260   & & P34867 & P14595 \\
3    & \object{HD 14956}  & B1.5\,Ia\tablefootmark{e} & Per OB1                   &  7.204$\pm$0.017   &    0.726$\pm$0.014   & FOCES      & 2005-09-25   & 2800               & 250   & & P40231 & P19317 \\
4    & \object{HD 91316}  & B1\,Iab\tablefootmark{d}  & Field                     &  3.847$\pm$0.014   & $-$0.142$\pm$0.007   & ESPaDOnS   & 2008-12-05   & 2$\times$320       & 920   & &   P19501 & R15529 \\
5    & \object{HD 114199} & B0.5\,IIe\tablefootmark{f}& Field                     &  9.474$\pm$0.008   &    0.296$\pm$0.011   & FEROS      & 2013-05-08   & 2$\times$1800      & 265   & & ...    & ...   \\
6    & \object{HD 152235} & B0.7\,Ia\tablefootmark{d} & NGC 6231                  &  6.329$\pm$0.021   &    0.510$\pm$0.016   & FEROS      & 2005-04-24   & 500\,+\,600        & 440   & & P16205 & R12470 \\
\hline
\end{tabular}
\tablefoot{
\tablefoottext{a}{\cite{Mermilliod97}}
\tablefoottext{b}{\cite{Melenaetal08}}
\tablefoottext{c}{revised here to: in the foreground of NGC~3603}
\tablefoottext{d}{Walborn's B-type standards \citep{GrCo09}}
\tablefoottext{e}{\cite{Lennonetal92}}
\tablefoottext{f}{revised here from B1\,Ia \citep{Houketal76}}
}}
\end{table*}

\section{Observational data}\label{section:observations}
As the spectral type B1 of Sher~25 lies slightly outside the range covered in Paper~I, an additional five supergiants
of types B1.5 to B0.7 were selected for comparison here. The quantitative analyses of high-quality spectra of B-type 
supergiants will thus be extended to hotter effective temperatures, based on the same homogeneous analysis methodology.

Phase~3 data on three individual spectra of Sher~25, and data for HD~114199 and HD~152235, as observed with the 
Fiberfed Extended Range Optical Spectrograph \citep[{FEROS},][]{Kauferetal99} on the Max-Planck-Gesellschaft/European 
Southern Observatory (ESO) 2.2\,m telescope at La Silla in Chile were downloaded from the ESO Science 
Portal\footnote{\rule{-0.25mm}{0mm}\url{https://archive.eso.org/scienceportal/home}}. They cover a useful wavelength range from about
3800 to 9200\,{\AA} at $R$\,=\,$\lambda / \Delta \lambda$\,$\approx$\,48\,000. The Sher~25 spectra, which were 
taken in one night, were co-added in order to increase the $S/N$.
A pipeline-reduced spectrum of HD~91316 ($\rho$\,Leo) as observed with the Echelle Spectro-Polari\-metric Device for the 
Observation of Stars \citep[{ESPaDOnS},][]{ManDon03} on the  3.6\,m Canada-France-Hawaii telescope (CFHT) at Mauna
Kea/Hawaii was downloaded from the CFHT Science Archive at the Canadian Astronomy Data 
Centre\footnote{\mbox{\rule{-0.25mm}{0mm}\url{https://www.cadc-ccda.hia-iha.nrc-cnrc.gc.ca/en/cfht}}}. It covers a wavelength range from 
about 3700 to 10\,500\,{\AA} at a resolving power of $R$\,=\,$\lambda / \Delta \lambda$\,$\approx$\,68\,000. 
The FEROS and ESPaDOnS spectra were  normalised by fitting a spline function through carefully selected continuum 
points. 

Finally, HD~13854 and HD~14956 were observed with the Fibre Optics Cassegrain Echelle Spectrograph 
\citep[{FOCES},][]{Pfeifferetal98} on the 2.2\,m telescope at the Calar Alto Observatory in Spain. The FOCES
spectra cover a wavelength range from 3860 to 9400\,{\AA} with $R$\,$\approx$\,40\,000. 
For FOCES the raw data needed to be reduced. Initially, a 
median filter was applied to the raw images to remove bad pixels and cosmics. The FOCES semi-automatic pipeline
\citep{Pfeifferetal98} was then used for the data reduction: subtraction of bias and darks, flatfielding, 
wavelength calibration based on Th-Ar exposures, rectification and merging  of the echelle orders. In a last step, 
all spectra were shifted into the laboratory rest frame via cross-correlation with appropriate synthetic spectra. 

Figure~\ref{fig:spect_lum_showcase_new} displays the spectra of the sample stars in three exemplary diagnostic wavelength windows, {\sc i})
around H$\delta$ with \ion{Si}{ii} and \ion{Si}{iv} lines, and several \ion{He}{i} and \ion{O}{ii} lines, 
{\sc ii}) the window around the \ion{Si}{iii} triplet to the \ion{He}{ii} $\lambda$4686\,{\AA} line and {\sc  iii}) on the H$\alpha$ 
emission line, with the neighbouring red \ion{C}{ii} doublet. The narrow features around H$\alpha$ are telluric H$_2$O 
lines.

Table~\ref{tab:spectra_info} summarises important observational information on the sample stars and provides an observing 
log. An internal ID number is given, the Henry-Draper catalogue designation, the spectral type and an OB association or 
open cluster membership is indicated. Then, Johnson $V$ magnitudes and the $B-V$ colours are given and the observing log 
contains information on the spectrograph, the observational date, exposure times, and the resulting $S/N$ 
of the final spectrum, measured around 5585\,{\AA}.

The star HD~114199, which is subject to a model atmosphere analysis 
for the first time, was classified as B1\,Ia in the literature \citep{Houketal76}. However, closer inspection of the 
spectrum finds significantly higher rotational velocity than in the other sample stars, broader Balmer lines and 
'double-horned' emission in H$\alpha$ (see Fig.~\ref{fig:spect_lum_showcase_new}). This is reminiscent of a Be star with an indistinct disk, seen under 
an intermediate inclination angle, instead of the emission arising from a stellar wind. By comparison with \object{HD 218376} 
(Cas~1, B0.5\,III) from the list of Walborn's B-type standard stars \citep{GrCo09} HD~114199 is reclassified as B0.5\,IIe here, 
because of its narrower Balmer lines and an otherwise very similar spectrum.

Several sources of (spectro-)photometric data were employed for the present work in addition to the Echelle 
spectra. Low-dispersion, large-aperture spectra taken with the International Ultraviolet Explorer 
(IUE, see also Table~\ref{tab:spectra_info}) were downloaded from the  Mikulski Archive 
for Space Telescopes (MAST\footnote{\url{https://archive.stsci.edu/iue/}}). 
Photometric measurements in the ultraviolet wavelength range include data from the 
Astronomical Netherlands Satellite \citep[ANS,][]{wesseliusetal82} and the Belgian/UK Ultraviolet Sky Survey 
Telescope \citep[S2/68,][]{Thompson95} on board the European Space Research Organisation (ESRO) TD1 satellite. For Sher~25, we adopted UV-photometry obtained with the Ultraviolet/Optical Telescope (UVOT) on board the Neil Gehrels Swift 
Observatory \citep{Yershov14}. Optical low-resolution spectra were provided by Gaia DR3. 
In addition, Johnson $UBV$ magnitudes \citep{Mermilliod97}, $UBVRI$ magnitudes \citep{Sung04}, $JHK$ magnitudes from the Two Micron All
Sky Survey \citep[2MASS,][]{Cutrietal03} and from \citet{HaEiMa08}, and Wide-Field Infrared Survey Explorer (WISE) photometry 
\citep{Cutrietal21} were adopted in the course of this work.

\section{Models and analysis methodology}\label{section:models}
The quantitative analysis of B-type supergiants requires consideration of deviations from local thermodynamic 
equilibrium (non-LTE effects). The hybrid non-LTE approach discussed in Paper~I as well as the same analysis 
technique were adopted for the present work, similar to other applications on e.g.~BA-type supergiants 
\citep{Przybillaetal06,SchPr08,FiPr12} or on the supergiants' progenitors, early B-type \citep{NiPr07,NiPr12,NiPr14} and late O-type 
main-sequence stars \citep{Aschenbrenneretal23}, or envelope-stripped massive B-type stars \citep{Irrgangetal22}.

\begin{table}
\caption{Model atoms for non-LTE calculations with {\sc Detail}.}             
\label{table:modelatoms}      
\centering                        
{\small
\begin{tabular}{llll}        
\hline\hline
Ion                         & Terms           & Transitions    & Reference \\ \hline
H                           & 25              & 300            & {[}1{]}   \\
He\,{\sc i/ii}              & 29+6/20         & 162/190        & {[}2{]}   \\
C\,{\sc ii/iii}             & 68/70           & 425/373        & {[}3{]}   \\
N\,{\sc ii}                 & 77              & 462            & {[}4{]}   \\
O\,{\sc i/ii/iii}           & 51/176+2/132+2  & 243/2559/1515  & {[}5{]}   \\
Ne\,{\sc i/ii}              & 153/78          & 952/992        & {[}6{]}   \\
Mg\,{\sc ii}                & 37              & 236            & {[}7{]}   \\
Al\,{\sc iii}               & 46+1            & 272            & {[}8{]}   \\
Si\,{\sc ii/iii/iv}         & 52+3/68+4/33+2  & 357/572/242    & {[}9{]}   \\
S\,{\sc ii/iii}             & 78/21           & 302/34         & {[}10{]}  \\
Fe\,{\sc iii/iv}            & 60+46/65+70     & 2446/2094      & {[}11{]}\\\hline
\end{tabular}
\tablefoot{Data for different ionisation stages are separated by a slash. If present, the number of superlevels is indicated after a plus sign.}
\tablebib{[1] \cite{PrBu04}; [2] \cite{przybilla05}; [3]~\cite{NiPr06,NiPr08}; [4] \cite{PrBu01}; [5] \cite{Przybillaetal00}, Przybilla \& Butler (in prep.); [6]~\cite{MoBu08}; [7] \cite{Przybillaetal01a}; [8] Przybilla (in prep.); [9] Przybilla \& Butler (in prep.); [10] \citet{Vranckenetal96}, updated; [11] \cite{Moreletal06}, updated.
}}
\end{table}

In brief, 
line-blanketed LTE model atmospheres were computed with {\sc Atlas9}  \citep{Kurucz93}, while non-LTE line-formation calculations were performed using 
extended and updated versions of {\sc Detail} and {\sc Surface} \citep{Giddings81,BuGi85}, adopting state-of the 
art model atoms. Information on the model atoms is summarised in Table~\ref{table:modelatoms}, which lists the 
number of (usually $LS$-coupled) terms (and superlevels) considered for the given ionisation stage and the number of 
radiative bound-bound transitions explicitly accounted for in the non-LTE calculations and references where a 
detailed description of the model atom can be found. The model grids described in Paper~I were extended to an 
effective temperature of 27\,000\,K. In order to compare the synthetic and observed spectra, 
the Spectral Plotting and Analysis Suite \citep[{\sc Spas},][]{Hirsch09} was used.

The atmospheric parameters were derived using multiple independent spectroscopic indicators simultaneously. The 
effective temperature $T_\mathrm{eff}$ and surface gravity $\log g$ were derived by
fitting the Stark-broadened Balmer lines and ionisation equilibria of \ion{He}{i/ii} and several metals, e.g. \ion{Si}{ii/iii/iv}. The microturbulent velocity $\xi$ was determined in the standard
way by demanding abundances to be independent of line equivalent widths. Projected rotational velocities $\varv 
\sin i$, macroturbulent velocities $\zeta$ and the elemental abundances for species $X$ relative to the hydrogen 
abundance, $\varepsilon(X)$\,=\,$\log(X/$H)\,+\,12, were determined from fits to individual line profiles. 
The helium abundance $y$ (by number) was derived by considering only the weakest helium lines, as the saturation 
of the stronger features restricts their sensitivity to abundance changes. 

In order to characterise interstellar reddening, both the total-to-selective extinction $R_V$\,=\,$A_V$/$E(B-V)$
and the colour excess $E(B-V)$ were determined, with $A_V$ being the interstellar extinction. {\sc Atlas} models 
of the spectral energy distribution (SED) were reddened using the mean extinction law of \citet{fitzpatrick99} in 
order to match the observations. An additional blackbody emitter was considered in the case of Sher~25 (see Sect.~\ref{section:discussion} and Appendix~\ref{appendix:A}).

Geneva evolutionary models for rotating stars \citep{Ekstroemetal12} were used to derive evolutionary masses 
$M_\mathrm{evol}$, which together with the surface gravity values allowed spectroscopic distances $d_\mathrm{spec}$ to the stars to be derived 
\citep[see Eq.~3 of][]{Wessmayeretal22}.
Bolometric corrections $B.C.$ were computed from the {\sc Atlas} fluxes. Absolute visual magnitudes $M_V$ were then derived from the apparent $V$ magnitudes and $d_\mathrm{spec}$-values, allowing for a determination of $M_\mathrm{bol}$ and the stellar luminosities $L$. Stellar radii $R$ were then constrained by combining the luminosities with $T_\mathrm{eff}$-values. A comparison with isochrones (from the evolutionary models) provided the evolutionary stellar ages $\tau_\mathrm{evol}$. Finally, consideration of Gaia DR3 parallaxes 
provided Gaia-based distances $d_\mathrm{Gaia}$, adopted as 'photogeometric distances' 
of \citet{Bailer-Jones_etal_2021}.

\begin{table*}
\caption{Stellar parameters of the sample stars.}
\label{tab:stellar_parameters}
\centering   
{\footnotesize
\setlength{\tabcolsep}{1mm}
\begin{tabular}{rlr@{\hspace{0.1mm}}rrcrrrrccrrrrrrrr}
\hline\hline
ID\#  & Object     &       & $T_{\mathrm{eff}}$ & $\log g$ & $y$ & $\xi$                & $\varv  \sin i$          & $\zeta$              & $R_V$ & $E\left(B-V\right)$ & $B.C.$  & $M_V$ & $M_\mathrm{bol}$  & $M_{\mathrm{evol}}$ & $R$         & $\log L/L_\sun$ & $\log \tau_\mathrm{evol}$ & $d_{\mathrm{spec}}$ & $d_{\mathrm{Gaia}}$\tablefootmark{a} \\ \cline{7-9}
     &        &       & kK                 & (cgs)      & & \multicolumn{3}{c}{$\mathrm{km\,s}^{-1}$} &       & mag                 & mag    & mag & mag   & $M_{\odot}$         & $R_{\odot}$ &  & yr & pc                  & pc                  \\ \hline
1                    & Sher 25                 &                      & 20.9             & 2.61                 & 0.117                   & 16                    & 60                   & 35                  & 3.40                 & 1.66               & $-1.920$ & $-7.05$ &  $-8.97$   & 25.5                 & 42                  & 5.48    & 6.86             & 5440                 & 5740                 \\
                     &                      & $\pm$                & 0.5                  & 0.06                 & 0.015                   & 2                    & 5                    & 5                    & 0.1                  & 0.03                 &   & 0.34 & 0.34 & 1.7                  & 7                    & 0.14       & 0.04         & 700                  & $^{800}_{420}$        \\
2                    & HD 13854                 &                      & 20.5                 & 2.62                 & 0.111                   & 15                    & 57                   & 47                   & 3.02                 & 0.52                 & $-1.871$  & $-6.91$ & $-8.78$ & 23.7                 & 40                   & 5.41    & 6.88             & 2320                 & 2140                 \\
                     &                      & $\pm$                & 0.3                  & 0.04                 & 0.010                   & 2                    & 6                    & 5                    & 0.1                  & 0.03                 &  & 0.21 & 0.22 & 1.0                  & 4                    & 0.09       & 0.03         & 200                  & $^{110}_{100}$        \\
3                    & HD 14956                 &                      & 19.2                 & 2.37                 & 0.127                   & 16                    & 49                   & 55                   & 2.69                 & 0.97                 & $-1.722$ & $-7.55$ & $-9.27$ & 27.2                 & 57                   & 5.60    & 6.81             & 2680                 & 2720                 \\
                     &                      & $\pm$                & 0.3                  & 0.04                 & 0.010                   & 2                    & 7                    & 5                    & 0.1                  & 0.03                 &   & 0.24 & 0.24 & 1.5                  & 7                    & 0.10       & 0.03         & 250                  & $^{150}_{130}$        \\
4                    & HD 91316                 &                      & 21.7             & 2.87                 & 0.092                   & 15                    & 43                   & 62                  & 2.81                 & 0.09                 & $-2.007$ & $-6.19$ & $-8.20$ & 19.9                 & 28                  & 5.18    & 6.96             & 900                 & 660                 \\
                     &                      & $\pm$                & 0.2                  & 0.04                 & 0.009                   & 2                    & 5                    & 5                    & 0.1                  & 0.03                 &   & 0.20 & 0.20 & 1.0                  & 3                    & 0.08       & 0.04         & 70                  & $^{170}_{140}$        \\
4a\tablefootmark{b,c} &                           &                      &                  &                      &                         &                       &                      &                     &                      &                      &   & $-5.51$ & $-7.52$                 & 10.6                 & 20                  & 4.90    & ...             & ...                 &                  \\
                     &                      & $\pm$                &                   &                  &                    &                     &                     &                     &                   &                  &  & 0.52 &  0.53                   & 5.3                  & 5                    & 0.21       & ...         & ...                  &         \\
5                    & HD 114199                 &                      & 25.6                 & 3.42                 & 0.119                   & 14                    & 135                   & 50                   & 3.21                 & 0.59                 & $-2.436$  & $-4.86$ & $-7.29$ & 15.7                 & 13                   & 4.81    & 7.03             & 3070                 & 2630                 \\
                     &                      & $\pm$                & 0.4                  & 0.05                 & 0.010                   & 2                    & 9                    & 10                    & 0.1                  & 0.03                 &   & 0.27 & 0.27 & 2.0                  & 2                    & 0.11       & 0.05         & 340                  & $^{100}_{100}$        \\
6                    & HD 152235                 &                      & 21.6              & 2.68                 & 0.085                   & 16                    & 59                   & 50                  & 3.00                 & 0.80                & $-2.006$   & $-6.89$ & $-8.89$ & 24.5                 &  38                  &  5.45
   & 6.87             & 1460
                 & 1620                 \\
                     &                      & $\pm$                & 0.3                  & 0.05                 & 0.008                   & 2                    & 4                    & 5                    & 0.1                  & 0.03                 &  & 0.24 & 0.25 &    1.3               &  4                   & 0.10       & 0.04         &  140                 & $^{130}_{90}$        \\
 \hline
\end{tabular}
\tablefoot{Uncertainties are 1$\sigma$-values, except where noted otherwise. 
\tablefoottext{a}{\cite{Gaia2016,Gaia2020} -- distances and uncertainties correspond to 'photogeometric 
distances' and associated $14^{\mathrm{th}}$ and $86^{\mathrm{th}}$ confidence percentiles \citep{Bailer-Jones_etal_2021}.}
\tablefoottext{b}{Alternative solution, adopting the Gaia distance for the calculation of the fundamental stellar parameters instead of the spectroscopic distance, starting from $M_V$.}
\tablefoottext{c}{HD~91316 is in fact a binary star with the $V$ magnitude difference of the two components in the range $\sim$1 to 1.5\,mag at similar $T_\mathrm{eff}$ as discussed in Sect.~\ref{section:individual}. As the Gaia parallax is likely affected by the binary orbital motion (indicated by a high Renormalised Unit Weight Error of 2.457) and in the absence of a second line system, a further determination of the parameters of the binary components is beyond the scope of the present paper.}
}}
\end{table*}

\section{Results}\label{section:results}

\subsection{Atmospheric and fundamental stellar parameters}\label{section:stellar_parameters}
We summarise the results of the analysis of the sample objects in Table~\ref{tab:stellar_parameters}, listing the parameters as follows: internal identification number, object name or HD-designation, effective temperature, surface gravity, surface helium abundance (by number), microturbulent, projected rotational and macroturbulent velocities, total-to-selective extinction parameter, colour excess, bolometric correction, absolute visual and bolometric magnitudes, evolutionary mass, radius, luminosity, evolutionary age, spectroscopic and Gaia DR3 distances, that is 'photogeometric' distance estimations \citep{Bailer-Jones_etal_2021}. The associated $1\sigma$ uncertainty intervals are listed in the line below the derived quantities.

The uncertainties of the derived atmospheric parameters are largely consistent with those reported in Paper I. Effective temperatures were determined with a typical relative accuracy of $\delta T_{\mathrm{eff}}$\,$\approx$\,1--3\% and surface gravities with $\Delta \log g$\,$\approx$\,0.05\,dex. For surface helium abundances, the uncertainties are larger on average, but are generally consistent with $\delta y$\,$\approx$\,10\%.

The microturbulent velocity parameter is limited in uncertainty to the size and steps of the grid used in the fitting process. Though our analysis considered variations on scales of 1\,km\,s$^{-1}$ in some cases, the general grid was set up with a step size of 2\,km\,s$^{-1}$, such that we adopt $\Delta \xi$\,$\approx$\,2\,km\,s$^{-1}$ as a conservative uncertainty margin. Convergence of {\sc Atlas} atmospheres permit microturbulence values to reach up to $\xi$\,$\le$\,17\,km\,s$^{-1}$ in a few cases, though a value of $\xi$\,$=$\,16\,km\,s$^{-1}$ may be regarded as an upper limit within our analysis. We wish to stress that sample stars corresponding to this value nevertheless show consistent results across the range of inspected lines, irrespective of strength, species and ionisation stage (see the discussion in Sect.~\ref{section:abundances_and_metallicity}). The relative uncertainty of the projected rotational velocity amounts to $\delta \varv \sin(i)$\,$\approx$\,5--15\%, while absolute uncertainties for the macroturbulence were generally estimated at a value of 5\,km\,s$^{-1}$ (due to its large rotational velocity and the resulting ambiguity, HD~114199 was assigned an absolute uncertainty in macroturbulence of $\Delta \zeta$\,=\,10\,km\,s$^{-1}$). Uncertainties in total-to-selective extinction $R_V$ and colour excess $E\left(B-V\right)$ depend on the wealth of constraining (spectro-)photometric data available for fitting. However, consistent error-margins of $\Delta R_V = 0.1$ and $\Delta E\left(B-V\right) = 0.03$ mag are used to reflect the typical scatter. For the absolute visual and absolute bolometric magnitudes the uncertainties span a range of $\Delta M_V \approx \Delta M_\mathrm{bol}$\,=\,0.20--0.34\,mag. 

To determine the evolutionary mass $M_{\mathrm{evol}}$, the sample objects' location on the spectroscopic Hertzsprung-Russell diagram (sHRD, see the upper panel of Fig.~\ref{fig:hrd}) were compared to a grid of evolution tracks by \cite{Ekstroemetal12}. Interpolation yielded the ZAMS mass $M_{\mathrm{ZAMS}}$ facilitating the derivation of the evolutionary mass by tracing mass loss along the model track. This procedure normally produces uncertainties of $\delta M_{\mathrm{ZAMS}}$\,=\,$\delta M_{\mathrm{evol}}$\,$\approx$\,5\%, except for stars very close to the end of the main-sequence, specifically HD~114199. As the location of this star is consistent with different evolutionary stages in multiple tracks, its relative uncertainty in $M_{\mathrm{evol}}$ and all derived parameters are correspondingly larger than the 'typical' values. Stellar radii are determined to typically within $\delta R$\,$\approx$\,10--17\% and luminosities to $\Delta \log L/L_{\odot}$\,$\approx$\,0.08--0.14\,dex. For the evolutionary age $\log \tau_{\mathrm{evol}}$ of the sample stars, uncertainties are about $\Delta \log \tau_{\mathrm{evol}}$\,$\approx$\,0.04\,dex. The derived spectroscopic distances commonly show relative uncertainties of $\delta d_{\mathrm{spec}}$\,$\approx$\,8--13\%, in accordance with the mean relative 
difference between the deduced values and those inferred from Gaia parallaxes.

\subsection{Comparison with previous analyses}\label{section:comparison_previous_analyses}
Except for HD~114199, every object in the current sample has been analysed in one or even several previous studies. To improve the  comparison across the entire B-type supergiant regime, we also include the cooler supergiants of Paper I. For the sake of brevity we refer to our previous work for a description of the methodologies of some of the comparison studies (i.e. \citealp[, 7 objects in common]{Fraser_etal_10}; \citealp[, 11 objects]{IACOBIII}; \citealp[, 2 objects]{MaPu08}; \citealp[, 4 objects]{Searle_etal_08}). Here, we summarise the additional studies:\\
{\sc i)} \cite{Crowther_etal_06} employed the non-LTE stellar atmosphere codes {\sc Tlusty} \citep{Hubeny88,HuLa95} and {\sc Cmfgen} \citep{HiMi98} for their analyses. They estimated the effective temperature by matching the intensities of silicon lines of consecutive ionisation stages. For B0--B2 stars as investigated here the silicon line \ion{Si}{iv} 4089\,{\AA} was compared to the  \ion{Si}{iii} 4552--4574 multiplet. The surface gravity was constrained by reproducing H$\gamma$. For microturbulence, a standard $\xi$\,=\,20\,km\,s$^{-1}$ was assumed initially and adapted to
values in the range of 10--40\,km\,s$^{-1}$ if the He and Si lines could not be fitted consistently. A uniform abundance ratio of He/H\,=\,0.2 by number was assumed throughout. We have five objects in common.\\
{\sc ii)} \cite{Smarttetal02} used {\sc Tlusty} to generate non-LTE, hydrostatic H+He model atmospheres. The effective temperature was determined by fitting the \ion{Si}{iv} 4116\,{\AA} line and the triplet at \ion{Si}{iii} 4813-4830\,{\AA}. An estimate of the surface gravity was established by a fit to H$\gamma$ and H$\delta$. The microturbulent velocity was found by demanding equal abundances of the silicon multiplet \ion{Si}{iii} 4552--4574\,. The projected rotational velocities were determined by convolving model line profiles of multiple metal lines with a rotational broadening function and comparing with observation until a match was produced. Three objects are in common.\\
{\sc iii)} \cite{Hendryetal08} re-analysed the spectra of Sher~25 described above \citep{Smarttetal02}, using more advanced stellar atmosphere codes for their re-examination. The following three codes were employed: a refined version of {\sc Tlusty} considering metal-line blanketing, the hydrodynamic line-blanketed non-LTE code {\sc Fastwind} \citep{Santolaya-Rey97,Pulsetal05} and {\sc Cmfgen}. One object (Sher~25) is common.

\begin{figure*}[ht]
\centering 
    \resizebox{0.65\textwidth}{!}{\includegraphics[width=\hsize]{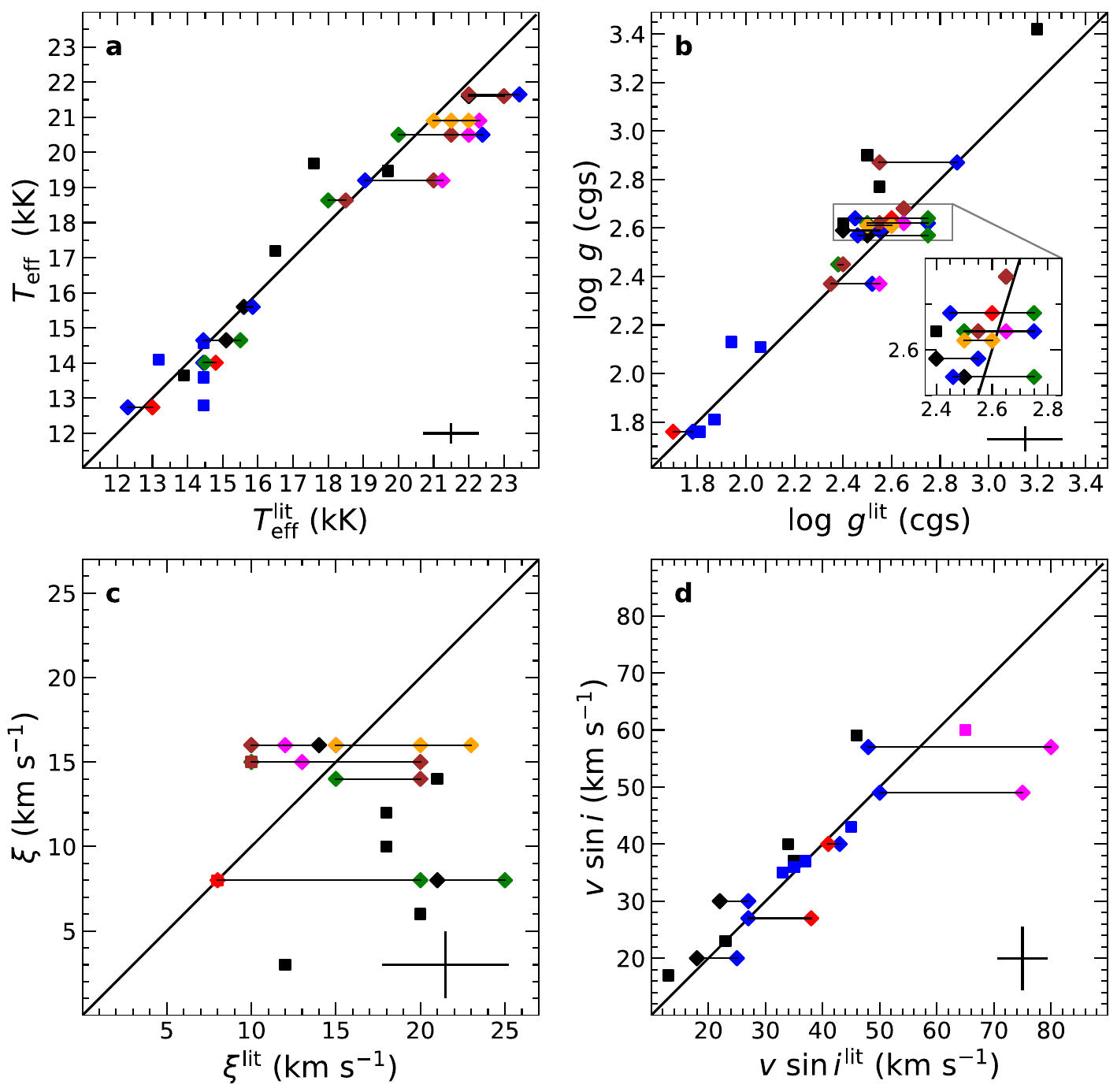}}
      \caption{Comparison of values for effective temperature $T_{\mathrm{eff}}$ (\textit{panel a}), surface gravity 
      $\log g$\, (\textit{panel b}), microturbulence $\xi$\, (\textit{panel c}), and projected rotational velocity 
      $\varv \sin i$ (\textit{panel d}) as derived in the present work and Paper~I with previous studies: 
\citet[][, black symbols]{Fraser_etal_10}, \citet[][, blue]{IACOBIII}, \citet[][, red]{MaPu08}, 
\citet[][, green]{Searle_etal_08}, \citet[][, brown]{Crowther_etal_06}, \citet[][, magenta]{Smarttetal02}, and 
\citet[][, orange]{Hendryetal08}. In cases in which an object is present in two or more studies the values are 
depicted by diamonds and connected with solid  black lines. For better visibility, an inset is added in panel~b. 
Mean error bars of the respective samples are indicated.}  
         \label{fig:param_comp}
\end{figure*}

Figure~\ref{fig:param_comp}, panel a, shows a comparison of this work's effective temperatures $T_{\mathrm{eff}}$ with those derived in the literature $T_{\mathrm{eff}}^{\mathrm{lit}}$. Large-scale, systematic offsets are absent across the set of discussed studies. When we look at the regime of the earlier supergiants investigated here, a small systematic offset towards higher temperatures may be noticed for some of the literature sets: for \citet{Smarttetal02} and \citet{Crowther_etal_06} the relative discrepancy is of the order of 8\% and 5\%, respectively. The \citet{Hendryetal08} analysis shows discrepancies of 0--5\%, while the two hottest objects in common with the \citet{IACOBIII} set are 9\% higher in temperature. A significant trend of this sort can be detected neither for the \citet{Fraser_etal_10} nor for the \citet{Searle_etal_08} studies. Naturally, these offsets must be put into perspective with the large scatter present between the compared works (i.e. between objects present in two or more literature studies, depicted as connected diamonds in Fig.~\ref{fig:param_comp}). In the case of HD~13854, the values range from $T_{\mathrm{eff}}$\,=\,20\,kK \citep{Searle_etal_08}, to 21.5\,kK  \citep{Crowther_etal_06}, 22\,kK \citep{Smarttetal02} and 22.4\,kK \citep{IACOBIII}. Similarly large ranges exist for all other compared objects. 

Values for the surface gravity are compared in Fig.~\ref{fig:param_comp}, panel b. In Paper I we discussed an emerging trend towards larger values of surface gravity derived using our methodology -- here, it is apparent that this effect is significantly diminished by addition of the earlier supergiants of this sample and the inclusion of further studies. However, a general offset persists for the set in common with \citet{Fraser_etal_10}, which shows lower surface gravities with $\Delta \log g$\,$\approx$\,0.2\,dex on average. Again, scatter among the studies is high: in the case of HD~13854, the estimates vary from $\log g$\,=\,2.5 \citep{Searle_etal_08}, to 2.55 \citep{Crowther_etal_06}, 2.65 \citep{Smarttetal02} and 2.75 \citep{IACOBIII}. At maximum, the differences amount to $\Delta\log g$\,$\approx$\,0.3\,dex.

\begin{table*}
\caption{Metal abundances $\varepsilon (X)$\,=\,$\log (X/\mathrm{H})+12$ and metallicity $Z$ (by mass) of the sample stars.}
\label{tab:abundances}
\centering   
{\small
\setlength{\tabcolsep}{1.5mm}
\begin{tabular}{lll@{\hspace{0.1mm}}llllllllllc}
\hline\hline
ID\#  & Object&          & C         & N         & O         & Ne        & Mg        & Al       & Si        & S         & Ar        & Fe  & $Z$      \\ \hline
1     & Sher 25   &       & 8.01 (5)  & 8.73 (23) & 8.64 (20) & 8.31 (7) & 7.42 (1)  & 6.35 (4) & 7.81: (8) & 7.47: (6) & ...  & 7.53 (8) & 0.016 \\
      &           & $\pm$ & 0.11      & 0.09      & 0.08      & 0.08      & ...      & 0.04     & 0.11      & 0.03      & ...      & 0.13      & 0.002 \\
2     & HD 13854   &       & 7.98 (6)  & 8.30 (34) & 8.52 (26) & 8.11 (5) & 7.25 (1)  & 6.17 (4) & 7.76: (10) & 7.33: (3) & 6.45 (1)  & 7.30 (12) & 0.010 \\
      &           & $\pm$ & 0.07      & 0.10      & 0.07      & 0.09      & ...      & 0.08     & 0.07      & 0.05      & ...      & 0.09      & 0.002 \\
3     & HD 14956   &       & 7.76 (7)  & 8.70 (27) & 8.53 (16) & 8.08 (8) & 7.37 (1)  & 6.30 (3) & 7.79: (9) & 7.34: (7) & ...  & 7.45 (7) & 0.013 \\
      &           & $\pm$ & 0.07      & 0.13      & 0.09      & 0.07      & ...      & 0.05     & 0.09      & 0.05      & ...      & 0.11      & 0.002 \\
4     & HD 91316~$^{a}$    &       & 7.94 (11)  & 8.35 (43) & 8.44 (30) & 8.06 (11) & 7.50 (4)  & 6.27 (4) & 7.60 (9) & 7.20 (4) & 6.57 (5) & 7.27 (17) & 0.010 \\
      &           & $\pm$ & 0.08      & 0.08      & 0.08      & 0.06      & 0.29      & 0.03     & 0.09      & 0.05      & 0.04      & 0.09      & 0.002 \\
5     & HD 114199   &       & 8.11 (4)  & 8.45 (19) & 8.53 (20) & 8.24 (4) & 7.46 (1)  & 6.57 (1) & 7.53 (6) & 7.05 (1) & ... & 7.61 (9) & 0.012 \\
      &           & $\pm$ & 0.08      & 0.07      & 0.07      & 0.06      & ...      & ...     & 0.06      & ...      & ...      & 0.09      & 0.002 \\
6     & HD 152235   &       & 8.30 (5)  & 7.85 (22) & 8.70 (27) & 8.15 (8) & 7.39 (1)  & 6.29 (4) & 7.72 (9) & 7.22 (3) & 6.62 (1) & 7.44 (7) & 0.013 \\
      &           & $\pm$ & 0.09      & 0.09      & 0.12      & 0.06      & ...      & 0.06     & 0.15      & 0.03      & ...      & 0.12      & 0.002 \\
\hline
& CAS~$^{b}$       &       & 8.35      & 7.79      & 8.76      & 8.09      & 7.56      & 6.30     & 7.50      & 7.14      & 6.50 & 7.52 & 0.014\\
&           & $\pm$ & 0.04      & 0.04      & 0.05      & 0.05      & 0.05      & 0.07     & 0.06   & 0.06      & 0.08 & 0.03 & 0.002\\ \hline
\end{tabular}
\tablefoot{Uncertainties are 1$\sigma$-values from the line-to-line scatter. Numbers in parentheses quantify number of
lines analysed. A colon is inserted for abundances with a suspected systematic overestimation (see Sect. \ref{section:abundances_and_metallicity} for details).  $^{(a)}$~Abundances are only indicative. The second continuum from the companion star needs to be 
considered for a proper abundance determination, which will result in higher abundances and metallicity. 
However, this is beyond the scope of the present work.~$^{(b)}$~cosmic abundance standard 
\citep[CAS,][]{NiPr12,Przybillaetal13}}}
\end{table*}  

Values for microturbulent velocity $\xi$ are shown in Fig.~\ref{fig:param_comp}, panel c. The comparison reveals a poor correlation between our results and previous studies, as well as large variation among the studies themselves, showing total spreads of up to $\Delta \xi$\,=\,12\,km\,s$^{-1}$. Adding the early supergiants to the picture, the variance is still large, but the correlation to our results improves.  

Finally, a comparison of $\varv \sin(i)$ values is presented in Fig.~\ref{fig:param_comp}, panel d. The trends are as observed in Paper I towards the higher velocities found here: very good accordance is apparent between our analysis and the set in common with \citet{IACOBIII}, while the results of \citet{Fraser_etal_10} indicate higher values by $\delta \varv \sin(i) \approx$\,20\% on average. The velocities reported by \citet{Smarttetal02} are much higher for two of the three common objects, but the line profiles were matched only by a rotational profile, ignoring macroturbulence unlike in 
the other studies. Nevertheless, we conclude in this context that overall the agreement between the more recent studies is good.

\subsection{Elemental abundances and stellar metallicity}\label{section:abundances_and_metallicity}
The analysis of elemental abundances of all metal species considered in our sample of stars is summarised in Table~\ref{tab:abundances}, featuring the mean abundance, uncertainty and number of analysed lines per species. The last column gives an estimate of the resulting metal mass fraction $Z$ (’metallicity’), calculated from the available mean abundances derived here. As the abundances derived in this work consider the ten most abundant metal species, these values should give a reliable estimate of the true mass fraction of metals. The error margins listed correspond to the $1\sigma$ standard deviation computed from the total set of analysed lines, giving equal weight to each line. These statistical error margins are comparable to those reported in Paper~I, though marginally larger, ranging from $\sim$0.05--0.15\,dex. The number of fitted lines per object and element is typically in the range of 5 to 10 and much larger in some cases, such that standard errors of the mean commonly amount to about 0.02\,dex. We omit uncertainty estimations for cases where only one line was suitable for fitting. For the metallicity, the $1\sigma$ uncertainty was estimated conservatively to be $0.002$ and adopted throughout the entire sample. 

The precise determination of atmospheric parameters, abundances and the careful treatment of macroturbulence and rotational broadening permit the production of global synthetic spectra capable of reproducing almost all spectral features found in the observed spectra, including blended lines excluded from our analysis. For Sher~25, a comparison between the observed spectrum and its global solution is discussed and shown in Appendix~\ref{appendix:C},
Figs.~\ref{fig:sher25_1} to \ref{fig:sher25_9}. A similarly close match between model and observation is also found for the other sample stars.

As our sample consists of objects scattered across the Galactic plane with differing distances $\rho$ from the Galactic centre (e.g. $\sim$7\,kpc for HD~152235 and over 10\,kpc for HD~14956) we cannot expect chemical homogeneity because of radial abundance gradients \citep[see e.g.][ for some more recent results]{daSilvaetal16,Bragancaetal19,ArellanoCordovaetal20}. Even so, it can be advantageous to discuss the overall picture in light of the cosmic abundance standard \citep[CAS,][]{NiPr12,Przybillaetal08b,Przybillaetal13}, which reflects the mean abundances derived for early B-stars in the chemically homogeneous solar neighbourhood (specified at the bottom of Table~\ref{tab:abundances}), providing a metallicity $Z$\,=0.014$\pm$0.002. Higher metallicities are expected for the sample stars in the inner Milky Way and lower metallicities well beyond the solar circle. Overall, agreement with this expectation is found for the sample stars within the error bars. However, Sher~25 appears to be metal-rich, whereas most of the other stars seem to be offset towards systematically lower metallicities. This is most pronounced for HD~91316 (ID\#4), which would result from the presence of significant second light in the binary system (see the discussion in Sect.~\ref{section:individual}). Second light from a circumstellar disk may also affect the abundance determination of HD~114199 (ID\#5). 

The metallicity is largely determined by the most abundant metals C, N, O and Ne, so that any systematic metallicity shifts are likely to originate from these. As the atmospheric parameters were derived achieving consistency simultaneously from several indicators and as they are in overall agreement with literature values, possible systematic abundance uncertainties may potentially stem from imperfections of the model atoms. However, the CNO mixing signatures (see Sect.~\ref{section:cno_signatures}) -- which are also an indicator for the quality of the atmospheric parameter analysis -- are also tightly matched, and the neon abundances are rather high, so no obvious source of the potential metallicity deficit can be identified. The \ion{C}{ii} $\lambda\lambda$4267 and 6578/82\,{\AA} lines are notoriously difficult to reproduce reliably (Appendix~\ref{appendix:C}) because of their complicated non-LTE line formation \citep{NiPr06,NiPr08}. They were consequently ignored for the carbon abundance determination, which relied mainly on the weaker \ion{C}{ii} lines (e.g. the quartet lines at $\lambda\lambda$5132--5151\,{\AA}) that are well reproduced in all cases. Nitrogen abundances were determined from the rich spectrum of \ion{N}{ii} lines and the oxygen abundances mostly from \ion{O}{ii} lines, which are around the turning point from major towards minor ionisation stage within the $T_\mathrm{eff}$-range investigated here. We note that small systematic underestimates of the abundances of mostly oxygen (and nitrogen when it is highly abundant) within the statistical 1$\sigma$-uncertainties suffice to bridge the gap to expected metallicity values.

\begin{figure*}[ht]
\centering 
    \resizebox{0.87\textwidth}{!}{\includegraphics[width=\hsize]{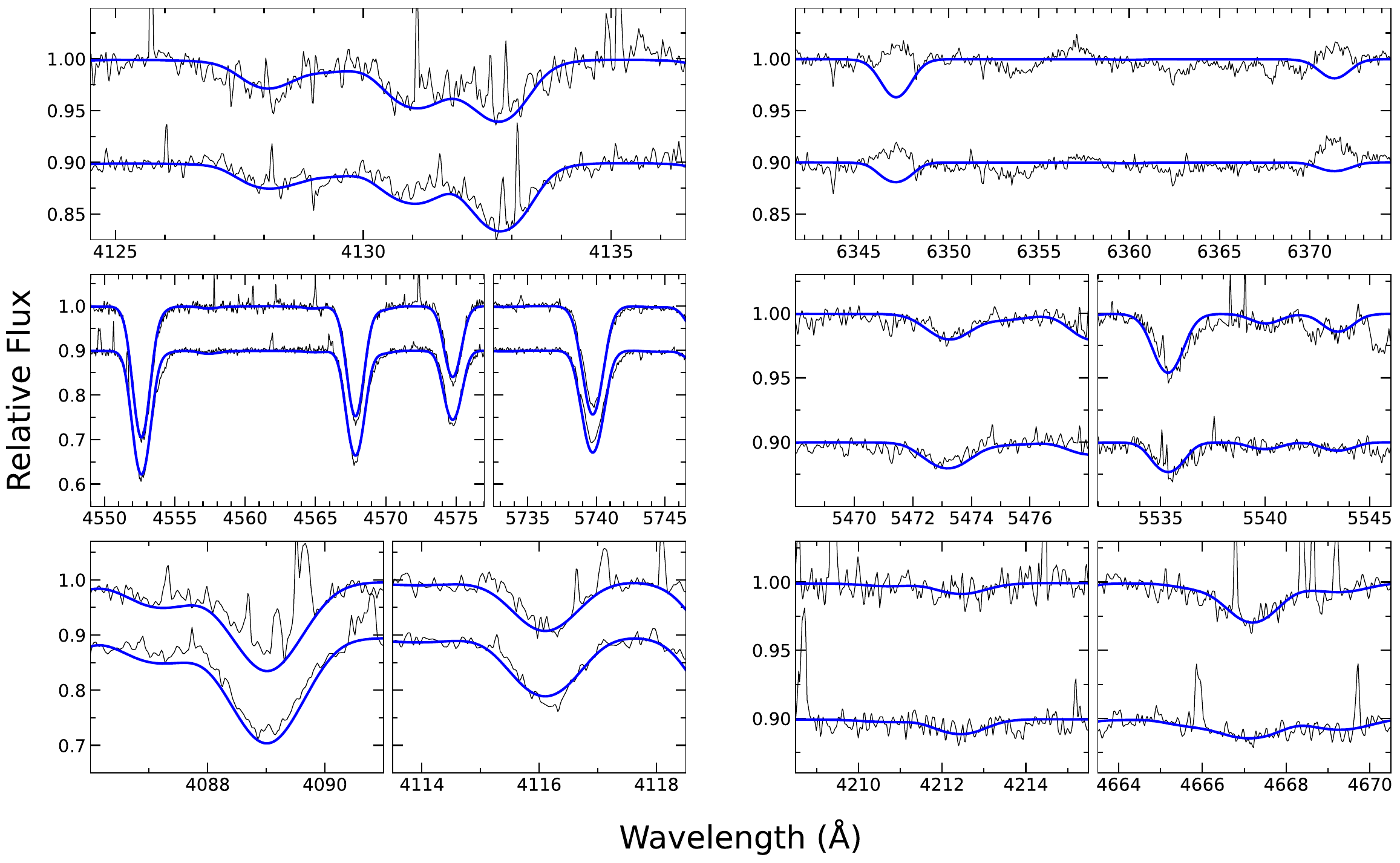}}
      \caption{Comparison of best-fitting models (blue) for observed spectra (black) of HD~13854 and HD~14956 (upper and lower lines in each sub-panel). The three rows show lines of different ionisation stages of silicon: Si\,{\sc ii} $\lambda$$\lambda$4128 and 4130\,{\AA} (\textit{upper left}), Si\,{\sc ii} $\lambda$$\lambda$6347 and 6371\,{\AA} (\textit{upper right}); the Si\,{\sc iii} triplet $\lambda$4552-4575\,{\AA} and Si\,{\sc iii} $\lambda$5739\,{\AA} (\textit{middle left}), Si\,{\sc iii} $\lambda$$\lambda$5473 and 5540\,{\AA} (\textit{middle right}); Si\,{\sc iv} $\lambda$$\lambda$4088 and 4116\,{\AA} (\textit{lower left}), Si\,{\sc iv} $\lambda$$\lambda$4212 and 4666\,{\AA} (\textit{lower right panel}).}  
         \label{fig:silicon_lines_sample}
\end{figure*}

Among the heavier elements, silicon and sulphur show higher abundances than expected for most stars, with maximum values reached in Sher~25. Lines of \ion{S}{ii} -- which is a minor ionisation stage at this $T_\mathrm{eff}$-range -- were analysed, as the \ion{S}{iii} implementation of the model atom of \citet{Vranckenetal96} is rather compact with 21 explicit non-LTE terms. Many of the energetically higher terms are also absent in the \ion{S}{ii} model, which may lead to an overpopulation of the existing terms relative to the true situation and potentially to an overestimated abundance determination. The presence of systematic effects cannot be excluded if only one ionisation stage is considered. A re-investigation using an improved sulphur model atom based on modern atomic data would be required to test this scenario, but this is beyond the scope of the present paper. 

The case of silicon is different, as lines from three ionisation stages are present in the observed spectra around the $T_\mathrm{eff}$-values investigated here, see Appendix~\ref{appendix:C} for the case of Sher~25 and Fig.~\ref{fig:silicon_lines_sample} for a selection of line fits in HD~13854 (ID\#2) and HD~14956 (ID\#3). Most of the lines from \ion{Si}{ii/iii/iv} -- weak and strong alike -- are reproduced simultaneously for the same abundance. This leaves little room for imperfections of the model atom, except for some details. Such are for example the \ion{Si}{ii} doublet lines $\lambda\lambda$6347 and 6371\,{\AA}, which are observed in emission but calculated in absorption. The lower level of the transition is radiatively coupled to the \ion{Si}{ii} ground state, which may make it sensitive to details of the overlap of the corresponding \ion{Si}{ii} resonance lines with spectral lines of other species\footnote{The \ion{He}{i} singlet problem \citep{Najarroetal06} in a restricted parameter range of late O-type stars may be seen as an analogue.}, which may drain the lower level's population, thus enabling the emission. We note that analogous calculations with the unified non-LTE model atmosphere code {\sc Fastwind} also fail to reproduce the observed emission (M.~Urbaneja, priv. comm.), implying that our neglect of sphericity and mass-outflow is not responsible for the failure. On the other hand, the \ion{Si}{iii} triplet $\lambda$4552-4575\,{\AA} is sensitive to mass-loss for very luminous supergiants \citep[see][]{Duftonetal05}. Overall, further investigations beyond the scope of the present work are required for the case of silicon to clarify the impact of modelling details on the line formation. The high silicon abundances derived here have to be viewed with caution until then.

\begin{figure}
\centering
\includegraphics[width=.95\hsize]{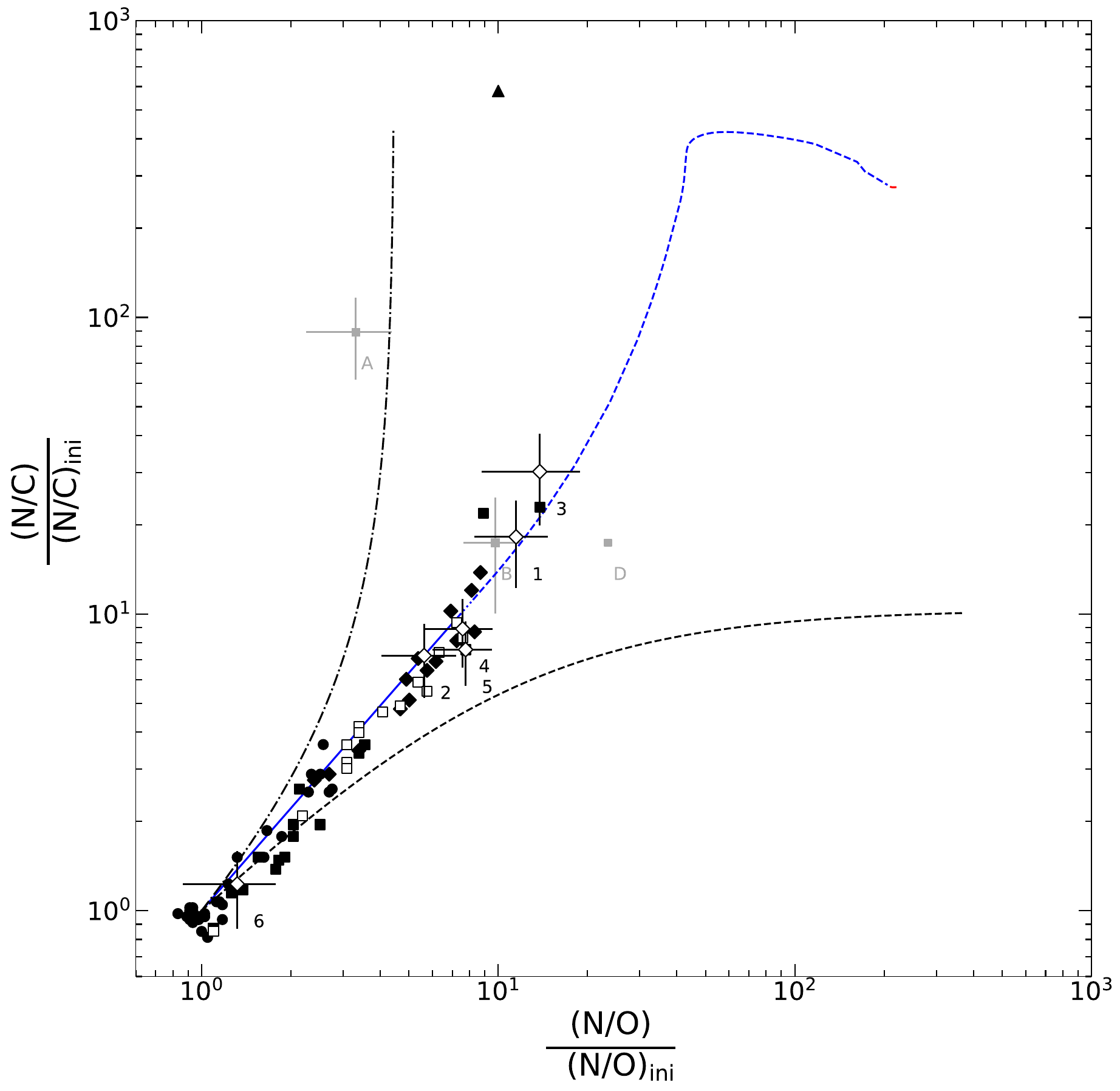}
\caption{Nitrogen-to-carbon ratio versus nitrogen-to-oxygen ratio, normalised to initial values. Objects from the present work are shown (open diamonds, marked by their ID\#) and from previous work employing an 
analysis methodology 
similar to the present one: B-type main-sequence stars \citep[][, black dots]{NiSi11,NiPr12}, BA-type supergiants 
\citep[][, black diamonds]{Przybillaetal10}, B-type supergiants (Paper I, open squares), late O-type main-sequence 
stars \citep[, black squares]{Aschenbrenneretal23} and the stripped CN-cycled core $\gamma$ Columbae \citep[, black 
triangle]{Irrgangetal22}. For comparison, the development of the surface CNO abundances is shown for
a 25\,$M_\sun$, $\Omega_{\mathrm{rot}}$\,=\,0.568\,$\Omega_{\mathrm{crit}}$ model by \citet{Ekstroemetal12}. The 
colour and style of the line depicts the different (main) evolution stages: ZAMS until TAMS (solid blue), further 
development until beginning of core He-burning (dotted blue), until core He exhaustion (dashed blue) and 
carbon-burning (red, at the very end of the track). The dashed black and dash-dotted lines depict the analytical 
boundaries for the ON- and CN-cycle, respectively \citep[cf.~Fig.~1 of][]{Maederetal14}. The grey squares are solutions for Sher~25 from the literature, see Sect.~\ref{section:discussion}. Their error bars are from standard errors of the mean CNO abundances, while error bars for results from the present work represent 1$\sigma$ standard deviations.}
\label{fig:cno_logplot}
\end{figure}

\subsection{Signatures of mixing with CNO-processed material}
\label{section:cno_signatures}
The atmospheres of rotating stars can be mixed with CNO-processed matter from the stellar core, facilitated by various physical effects such as meridional circulation or shear-mixing as a consequence of differential rotation \citep[e.g][]{MaMe12,Langer12}. The mixing may also be impacted by the presence of magnetic fields. The nitrogen-to-carbon (N/C) and nitrogen-to-oxygen (N/O) abundance ratios sensitively probe the degree of mixing with nuclear-processed material. Tracking stars in a N/C versus N/O diagram \citep[cf.~Fig.~5 of][]{Przybillaetal10} can help to gain insight into the evolutionary status of the examined stars. Moreover, as the graph shows only minor dependence on initial stellar masses, rotational velocities and the details of the mixing mechanisms for small relative enrichment (i.e. by a factor $\sim$4 over the initial N/O), it can serve to assess the overall quality of the observational results \citep{Maederetal14}.

Figure~\ref{fig:cno_logplot} shows the ratios of surface abundances for carbon, nitrogen and oxygen (normalised to 'initial' CAS values, see Table~\ref{tab:abundances}) for this work's sample stars and a collection of 76 objects analysed with a similar methodology to the one employed here, and three literature results for Sher~25 (see Table~\ref{table:parameters_sher25} and Sect.~\ref{section:discussion} for a discussion). The diagram also shows the limiting analytical solutions for the CNO-cycle: assuming constant (initial) oxygen abundance in the CN-cycle leads to the upper (almost vertical) boundary function, while the assumption of constant (equilibrium) carbon abundance in the ON-cycle leads to the lower (horizontal) solution. Predictions from stellar evolution calculations generally fall in between these limits, exemplified here by the model track for a rotating 25 $M_{\odot}$ star by \citet{Ekstroemetal12}. 

It is apparent that the bulk of analysed stars show mixing ratios of less than a factor $\sim$10 in both N/C and N/O above initial values, constrained tightly on the predicted evolutionary pathway. Four of the present sample stars also fall into this category. While HD~13854 (ID\#2) and HD~152235 (ID\#6) share very similar spectroscopic and fundamental parameters, the former shows mixing ratios typical for a supergiant at average rotation, whereas the latter features strikingly low CNO enhancement. Normal ratios, compatible with mixing having occurred on the main sequence, are also found for the rapidly rotating Be star HD~114199 (ID\#5) and possibly HD~91316 (ID\#4). 

Figure~\ref{fig:cno_logplot} also shows stars with significantly more enriched atmospheres (N/C\,$\gtrsim$\,20). Besides the extreme case of the stripped CN-cycled core $\gamma$~Columbae \citep{Irrgangetal22} towards the top of the figure and two ON-stars with binary mass-overflow history \citep[HD~14633 and HD~201345, described by][]{Aschenbrenneretal23} we also find two stars from the present sample as highly enriched, Sher~25 (ID\#1) and HD~14956 (ID\#3). While it is in principle feasible to produce such large enhancement for massive, rapidly rotating stars, a potential binary interaction with mass exchange has to be taken into consideration. In all cases, a detailed study of each individual star's unique characteristics aids in understanding the measured CNO signature (see Sect.~\ref{section:individual} for a discussion of each of our comparison stars and Sect.~\ref{section:discussion} for an in-depth account of the status of Sher~25). 

\begin{figure}
\centering
\includegraphics[width=.9\hsize]{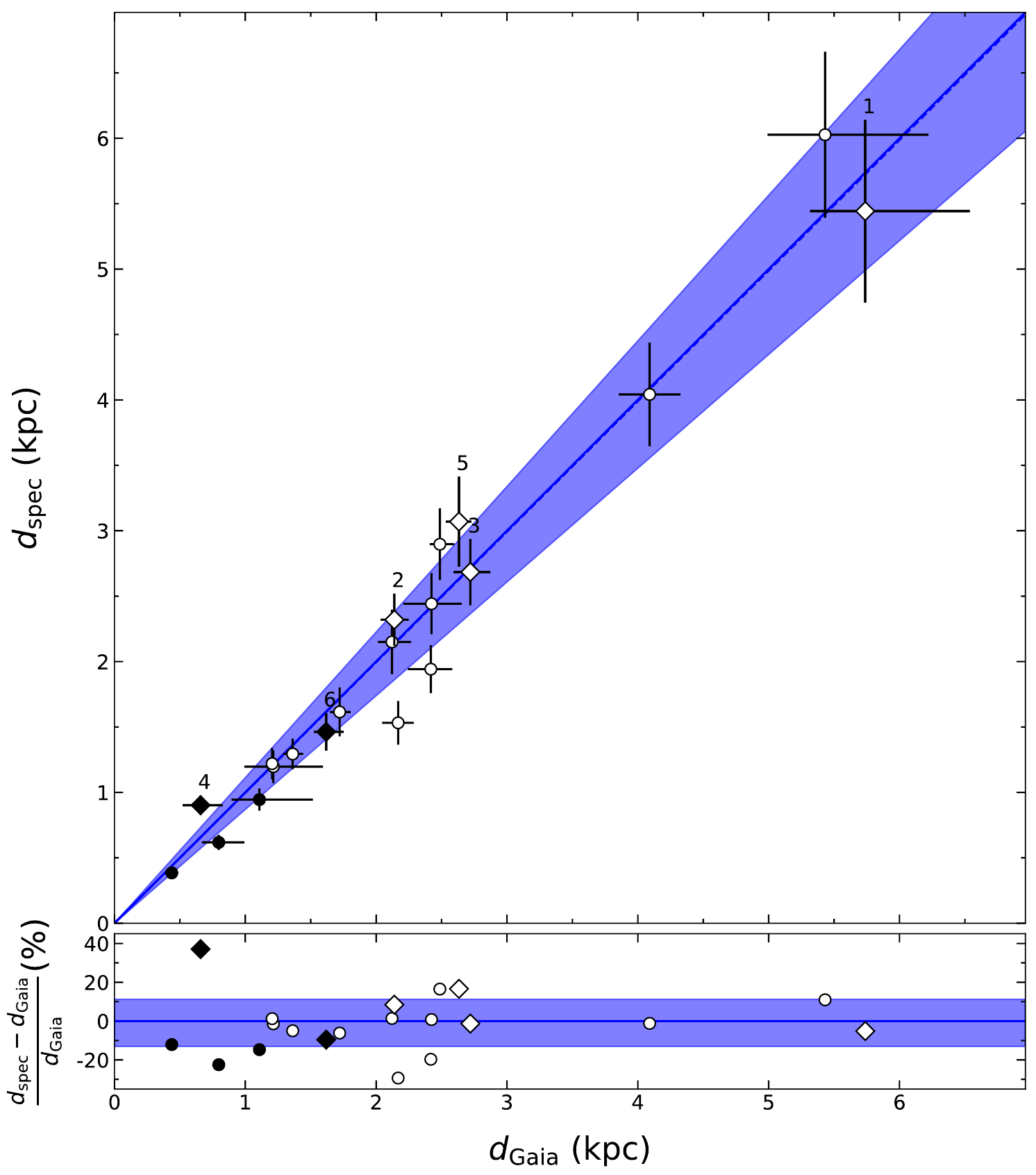}
      \caption{Comparison of derived spectroscopic distances and distances based on Gaia EDR3 parallaxes
      (\textit{upper panel}) and their relative differences (\textit{lower panel}). Diamonds represent the objects
      analysed in this work, while dots correspond to those of Paper I -- filled symbols are used to depict objects 
      with RUWE values $>$1.3. The solid blue lines depict equivalence, while the dashed line shows the best 
      linear fit to the data. The shaded area marks the region of $1\sigma$ standard deviation from the mean. In the 
      fit, only data with good RUWE values were employed, that is the open symbols.}
\label{fig:distance_comparison}
\end{figure}

\subsection{Spectroscopic distances}
The comparison of spectroscopic distances (determined in analogy to Paper~I) with those inferred from Gaia EDR3 parallaxes \citep[photogeometric distances of][]{Bailer-Jones_etal_2021} is shown in the upper panel of Fig.~\ref{fig:distance_comparison}, while the lower panel shows the relative differences. The figure also depicts stars discussed in Paper~I, providing a comprehensive overview to the overall quality of the correspondence between measurements across the entire domain of B-type supergiants. For the sub-sample of objects which can be compared reliably, that is, those with small renormalised unit weight error (RUWE\,$<$\,1.3), the relative differences show a mean offset of $\mu_s$\,=\,$-$1\% with a sample standard deviation of $\sigma_s$\,=\,12\%. The hot supergiants analysed in this work spread over a range in distance from 1 to about 5 kpc (Sher~25 being the most distant star of the sample), resembling the range of values of the entire sample. 

Evidently, the relationship shows no noticeable deviations either on the near or the far end of the distance scale. For IDs\#1, 2, 3, and 6, the accordance is good, with relative differences to the parallactic distance less than 10\%. Specifically, for HD~13854 (ID\#2) the spectroscopic distance is further corroborated by the star's membership in the open cluster \object{NGC 869} \citep[$d_{\mathrm{NGC869}}\approx2.3$ kpc, ][]{Currieetal2010}. Notable deviations exist only for two of the sample stars: for HD~91316 we derive a distance of 900$\pm$70\,pc, barely compatible with its parallax-based estimate of $660^{+170}_{-140}$\,pc (at a RUWE of  2.46). Even worse agreement is achieved with the star's Hipparcos distance of $1670^{+710}_{-380}$\,pc \citep{vanLeeuwen07a}. For HD~114199 the deviation is less extreme (but considerable) with our estimate of 3070$\pm$340\,pc being about 17\% above the Gaia value of 2630$\pm$100\,pc (RUWE\,$=$\,0.786). In the case of both of these objects the broader picture of their deduced characteristics hint at peculiarities in their evolutionary history, which are able to explain the observed discrepancy in distances, and we refer to Sect.~\ref{section:individual} for an in-depth discussion. Excluding them, it is clear that the resulting distances in the present work maintain the overall good agreement reported in Paper~I.   

\begin{figure}
\centering
\includegraphics[width=.87\hsize]{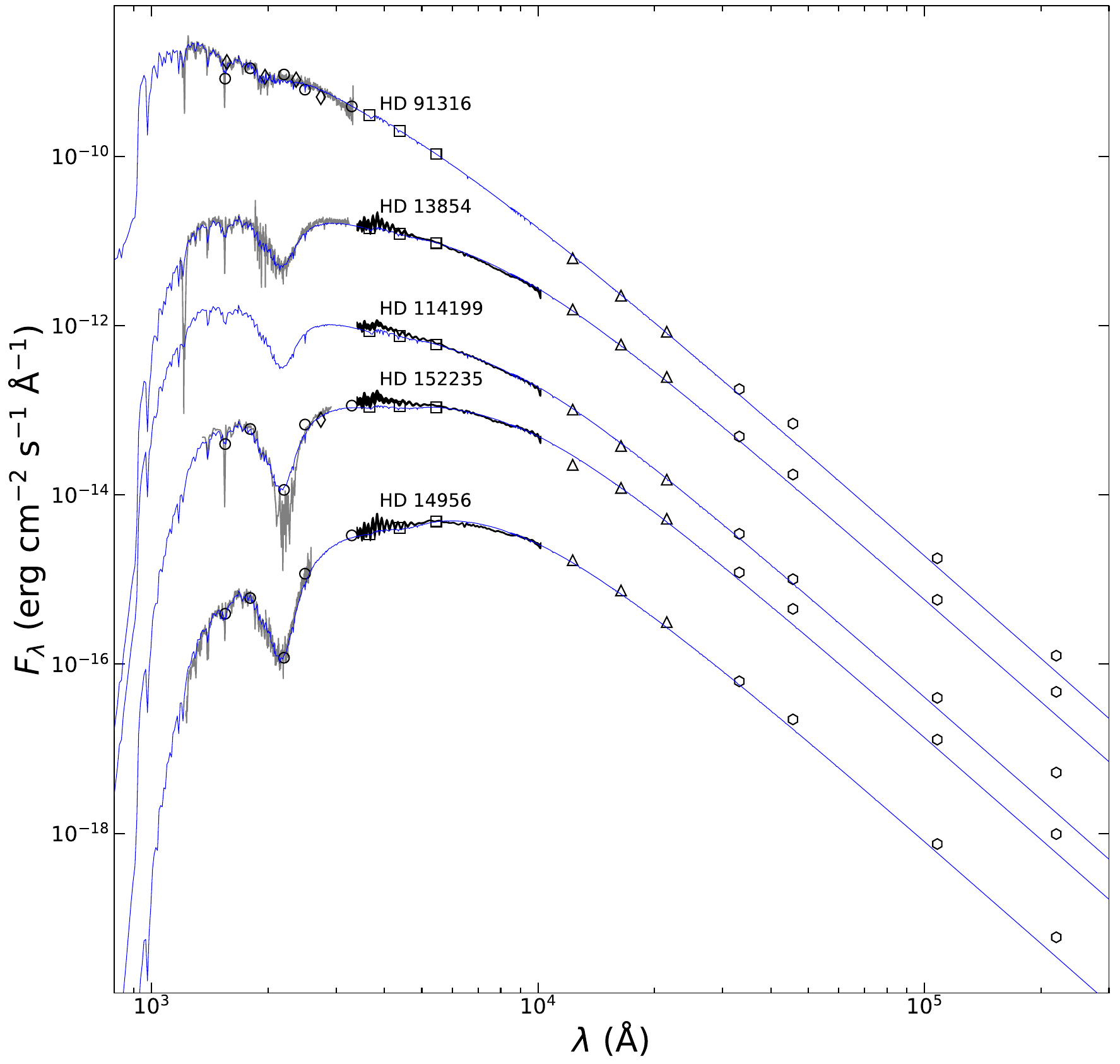}
      \caption{Spectral energy distributions of  the sample stars. {\sc Atlas9}-SEDs, normalised in $V$ and reddened 
      according to values from Table~\ref{tab:stellar_parameters} (blue lines) are compared  to IUE and Gaia spectrophotometry (grey and black lines, respectively) 
      and photometric data in various wavelength bands: ANS (circles), TD1 (diamonds),  Johnson (squares), 
      2MASS (triangles), and ALLWISE data (hexagons). For better visibility, the SEDs and photometry for HD~152235 and HD~14956 were shifted by $-2$ and $-3$\,dex, respectively.}
\label{fig:sed_plot}
\end{figure}

\subsection{Characterisation of the ISM sight lines}
\label{section:reddening_law}
While the methodological approach remains identical to the one described in Paper I, the determination of the reddening law for this work's sample made use of additional data, as summarised in Sect.~\ref{section:observations}. Here, the ISM sight lines towards the comparison stars will be discussed. We refer to Sect.~\ref{section:discussion} for the case of Sher~25, where additional constraints will be considered.

Figure~\ref{fig:sed_plot} shows the data and best-fitting model SEDs for the comparison stars, ordered by increasing colour excess $E(B-V)$ from top to bottom. For most objects, the model SED can explain all available (spectro-)photometric data sufficiently, though some ambiguities and mismatches must be explained. We see that for all objects the WISE band W4, and to a lesser extent also W2, show significant excess flux. While it is possible to reproduce those features assuming dust emission lines or, as in the case of HD~114199, a black body contribution from an indistinct disk around the star, these assumptions barely change the resulting reddening parameters. We consequently omit these data-points in the fitting process. In the case of HD~152235 the IUE spectrum shows excess absorption in the $\lambda$2175\,{\AA} extinction bump, not present in the ANS data. To fit the former would require the employment of an anomalous reddening law, while the latter can be consistently fit with all other measured data -- we therefore prefer the ANS data over the IUE spectrum (which is very noisy at these low flux levels). We also note that while the Gaia spectrophotometry is reproduced well overall, some deviations occur at the borders of the coverage. At the high wavelength limit a steep drop of the flux is observed at a few wavelength points, likely an issue from the calibration. In analogy, the Gaia spectrophotometry also overestimates the fluxes between the $U$ and $B$ bands, a known issue described in Sect.~8.2 of \citet{Montegriffoetal22},  see also their Fig.~38. On the other hand, the Johnson and ANS photometry and the IUE spectrophotometry can be nicely matched by the reddened models.

The derived values for $R_V$ and $E(B-V)$ of the comparison stars are comparable in range to the ones of Paper~I with both low and high values realised (see Table~\ref{tab:stellar_parameters}). For $R_V$, they vary between 2.7 to 3.2, while for the colour excess values range from about $E(B-V)$\,=\,0.1 to 0.8\,mag.

\begin{figure}
\centering
\includegraphics[width=0.8\hsize]{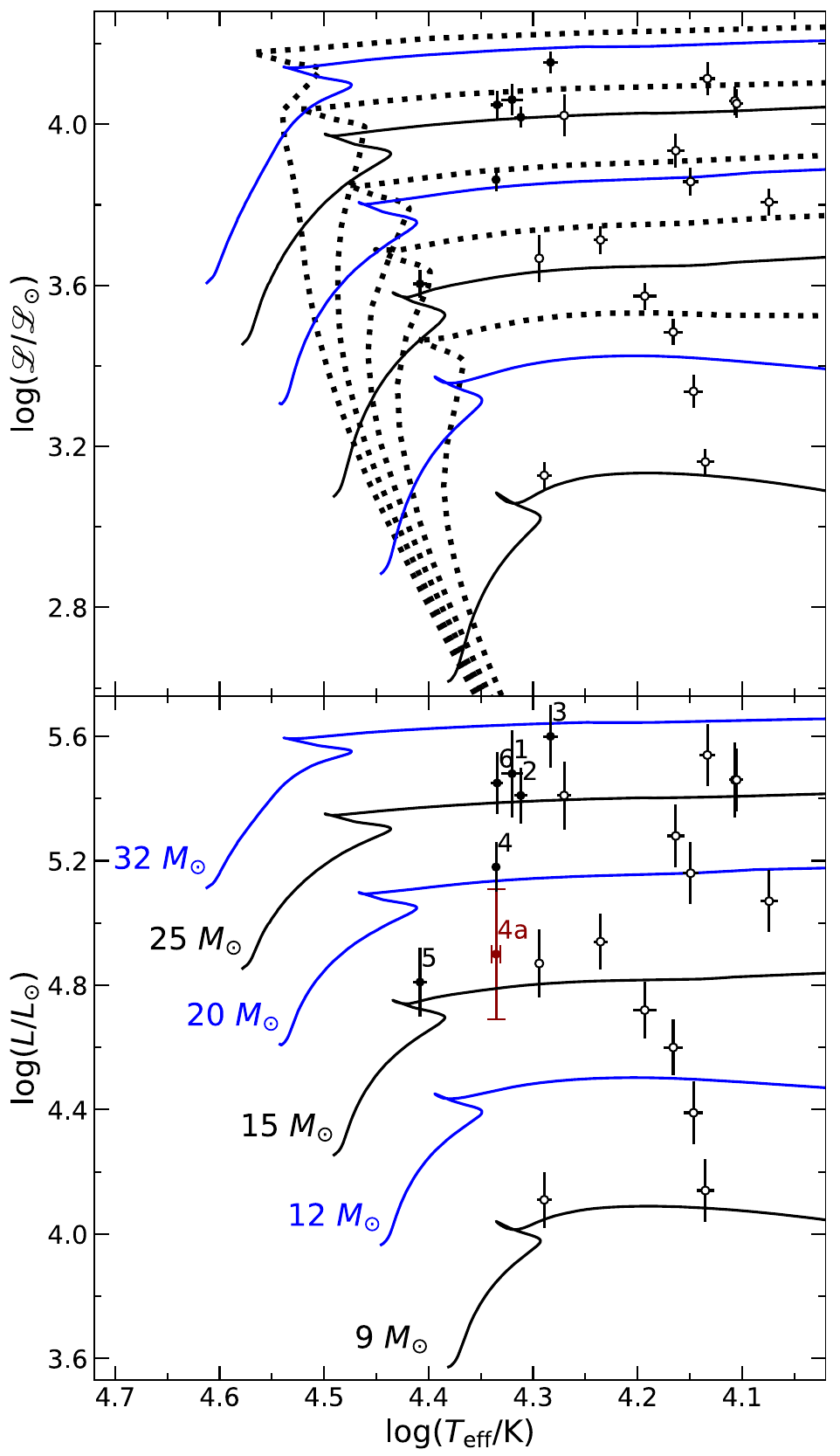}
\caption{Location of the sample objects (black dots with error bars) and B-type supergiants analysed in Paper~I 
(open symbols) in two diagnostic diagrams, the sHRD (\textit{upper panel}) and the HRD (\textit{lower panel}).
For comparison, loci of evolution tracks for stars rotating at 
$\Omega_{\mathrm{rot}}$\,=\,0.568\,$\Omega_{\mathrm{crit}}$ \citep{Ekstroemetal12} are indicated for various ZAMS-masses. Isochrones for the model grid, corresponding to ages of $\log \tau_\mathrm{evol} \in \{6.75, 
6.85, 6.95, 7.05, 7.20\}$ are depicted as dotted lines in the upper panel (increasing in age from top to bottom). For HD~91316 the luminosity as derived from the parallactic distance (ID\#4a in Table~\ref{tab:stellar_parameters}) is depicted with a red symbol and marked accordingly.  
Error bars indicate 1$\sigma$ uncertainty ranges.}
\label{fig:hrd}
\end{figure}

\begin{figure*}
\sidecaption
\includegraphics[width=12cm]{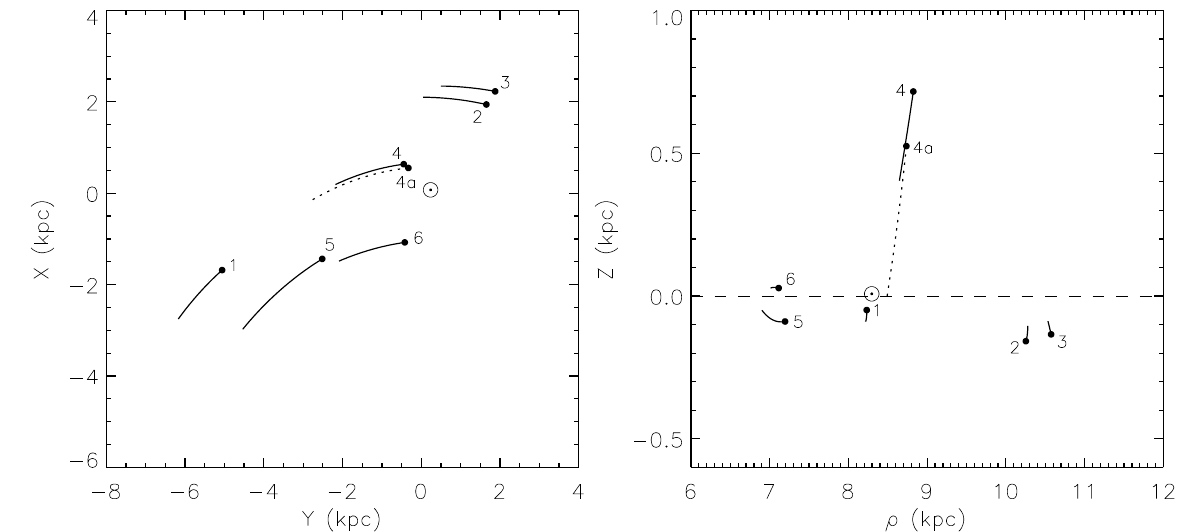}
\caption{Visualisation of the kinematics of the sample stars in the Galactic potential. Galactic Cartesian 
coordinates $XYZ$ are employed, with the origin shifted to the position of the Sun, adopting its galactocentric 
distance from the \citet{GravityCol1aboration19}. \textit{Left panel:} galactic plane projection. \textit{Right 
panel:} meridional projection. $\rho$ is the galactocentric distance and the Galactic mid-plane is indicated by the 
dashed line. Black dots mark the current positions of the stars, the curves show the trajectory calculated backwards 
in time according to the evolutionary age (see Table~\ref{tab:stellar_parameters}). The alternative solution ID\#4a 
(\textit{dotted line}) was calculated backwards until the Galactic mid-plane was reached, about 13\,Myr ago.}
\label{fig:kinematics}
\end{figure*}

\subsection{Evolutionary status}\label{section:evol_status}
Tracks of stellar evolution models can be used to derive the current evolutionary status of the sample objects. For this purpose we utilise two related diagnostic diagrams in Fig.~\ref{fig:hrd}: the spectroscopic HRD \citep[sHRD, $\log(\mathscr{L}/\mathscr{L}_{\odot})$ versus $\log T_\mathrm{eff}$, introduced by][]{LaKu14} and the HRD ($\log L/L_\odot$ versus $\log T_\mathrm{eff}$). While the stars' positions in the sHRD are determined by the spectroscopic solution alone ($\mathscr{L} = T^4_\mathrm{eff}/g$), the loci in the 'classic' HRD also rely on the objects' distance and reddening law. In addition to the present sample stars the cooler B-type supergiants from Paper~I are also shown in Fig.~\ref{fig:hrd}\footnote{Note that in the analogous Fig.~15 of Paper~I the uncertainties for those sample stars (marked by open symbols here) in the sHRD were also adopted for the HRD, which we correct here.}, together with evolutionary tracks and isochrones for rotating stars \citep{Ekstroemetal12}. The present sample stars, with half of them belonging to luminosity class Iab, are on average more massive and more luminous than those of Paper~I (about half belong to luminosity class Ia), reflecting the existence of much more luminous supergiants than discussed here on the hot side of the Humphreys-Davidson limit \citep{HuDa79}. 

Our present sample stars had ZAMS masses between about 15 and 30\,$M_{\sun}$ (but see also the discussion of HD~91316 (ID\#4/4a) in Sect.~\ref{section:individual}). The stellar ages range between about 6.5 to 11\,Myr, again with the exception of HD~91316. It has to be stated that the masses and ages were derived assuming that the rotation rates employed for the computation of the evolutionary tracks and isochrones are representative on average for the sample. Systematic shifts in mass and age result if the initial rotational velocities had other values, but we expect them to be covered by our uncertainties in most cases. The projected rotational velocities of the sample stars are slightly higher than in Paper~I, in agreement with predictions \citep{Ekstroemetal12} for slightly earlier evolutionary stages. Only HD~114199 rotates significantly faster, in accordance with its earlier evolutionary stage and its Be star nature. Another issue for the comparison between observation and models can be metallicity. The value of $Z$\,=\,0.014 employed by \citet{Ekstroemetal12} is representative for the solar neighbourhood, but stars at lower or larger galactocentric radius will experience the effects of abundance gradients. One can therefore expect metallicities to vary a bit among the sample stars (see the discussion in Sect.~\ref{section:abundances_and_metallicity}). However, the $Z$\,=\,0.014 tracks can still be viewed as representative because the closest published Geneva tracks for higher and lower metallicity are for $Z$\,=\,0.020 \citep{Yusofetal22} and 0.006 \citep{Eggenbergeretal21}, that is very distant in metallicity space. Moreover, the evolutionary tracks remain similar throughout this metallicity range \citep[see e.g.][, their Fig. 5]{Yusofetal22}.
 
The overall good agreement of the stars' positions in both the sHRD and HRD relative to the evolutionary tracks indicates that the sample supergiants are on their first crossing of the HRD towards the red. Otherwise, the high mass-loss experienced during the RSG phase would yield different positions, as also likely in the case of binary evolution -- as for HD~91316.

\subsection{Kinematics}\label{sect:kinematics}
It may be useful to consider trajectories in the Galactic potential for some stars to further constrain their 
evolutionary status from flight times. The Galactic potential as described by \citet{AlSa91} together and the code 
of \citet{OdBr92} were employed to calculate the Galactic orbits of the sample stars. Coordinates ($\alpha$, 
$\delta$) and proper motions ($\mu_\alpha$, $\mu_\delta$) in right ascension and declination and radial velocities 
for HD~13854 (ID\#2), HD~14956 (ID\#3) and HD~91316 (ID\#4/4a) were adopted from Gaia DR3 \citet{GaiaDR3}, 
spectroscopic distances (and $d_\mathrm{Gaia}$ for solution ID\#4a) from Table~\ref{tab:stellar_parameters} and 
radial velocities from \citet{Hendryetal08} for Sher~25 (ID\#1)\footnote{Systematic velocity, as the star is known to 
show radial velocity variations, possibly due to pulsations \citep{Tayloretal14}.}, from \citet{Gontcharov06} for 
HD~152235 (ID\#6) and a value of $-$45.6\,km\,s$^{-1}$ for HD~114199 (ID\#5) as determined from the observed 
spectrum. Figure~\ref{fig:kinematics} shows the resulting stellar trajectories in the Galactic plane and the 
meridional projection. Stars IDs\#1, 5 and 6 are consequently located in the Carina-Sagittarius spiral arm and stars 
IDs\#2 and 3 in the Perseus spiral arm. For all sample stars the 
kinematics is dominated by the rotation around the Galaxtic centre (at Galactic Cartesian coordinates $X$\,=
\,$-$8.178\,kpc, $Y$\,=\,0\,kpc, $Z$\,=\,0\,kpc in this visualisation), except for HD~91316 (ID\#4/4a). The star 
shows a clear runaway movement from its origin in the local spiral arm perpendicular to the Galactic plane, which 
explains its high galactic latitude. The implications for the evolutionary scenario of HD~91316 will be discussed in the next Section.

\section{Summary of individual comparison stars}\label{section:individual}
\paragraph{HD~13854 (ID\#2).} This is one of the supergiant members of the open cluster \object{NGC 869} (h Per) in 
the Per~OB1 association. It shows a CNO mixing signature that one may expect for a 'typical' supergiant of its mass 
at average rotation, see Fig.~\ref{fig:cno_logplot}. Its metallicity is lower than the standard value in the solar 
neighbourhood, as is expected from its position farther out in the Galactic disk. Other than this, it is 
the closest analogue to Sher~25 among all the sample stars discussed here because of its very similar atmospheric 
and fundamental stellar parameters.\\[-8mm]
\paragraph{HD~14956 (ID\#3).}
The star is a member of the Per~OB1 association. It is the most massive star of the entire sample, including the 
objects from Paper~I. HD~14956 shows the highest degree of CNO-mixing in Fig.~\ref{fig:cno_logplot}, except for the
stripped-core star $\gamma$~Col \citep{Irrgangetal22}. The position in the N/O-N/C diagram in conjunction with only a 
mild enrichment of helium could be reached already after termination of the main-sequence phase, if such a massive 
star was rotating somewhat faster than average \citep[cf. the evolutionary tracks of][]{Ekstroemetal12}. Predictions 
for a post-RSG scenario would locate the star close to the turning point immediately preceding the Wolf-Rayet phase
in Fig.~\ref{fig:cno_logplot}, with the CNO abundances dominated by convective dredge-up, irrespective as to whether
the star was rotating or not. Values of $y$\,$\approx$\,0.2 and a surface gravity lower by a factor two would be 
expected, which are not observed. We note, however, that the position of HD~14956 in Fig.~\ref{fig:cno_logplot}
is similar to that reached by the two ON stars \object{HD 14633} and \object{HD 201345} discussed by 
\citet{Aschenbrenneretal23}, which have obtained their high degree of CNO mixing by mass accretion in a binary 
scenario. HD~14956 was identified as a SB1 system 
with a period of $\sim$175\,days by \citet{AbLe73}, but see also \citet{deBurgosetal20}.\\[-8mm]
\paragraph{HD~91316 ($\rho$~Leo, ID\#4).}
The star is one of the few supergiants at high galactic latitude ($\ell$\,$\approx$\,$+53$\degr), a position reached 
as a runaway star. Figure~\ref{fig:kinematics} shows its trajectory in the Galactic potential. Solution \#4, adopting 
the spectroscopic distance, is obviously unable to trace the star back to a star-forming region close to the Galactic 
mid-plane -- where massive stars are born --  within the lifetime of the deduced $\sim$20\,$M_\odot$ star.
The Gaia DR3 distance is, on the other hand, significantly shorter, leading to a lower luminosity and therefore to a 
smaller mass and longer lifetime (solution \#4a). This may suffice to bring the trajectory back to the galactic mid-
plane within 13\,Myr, which may be compatible with a dynamical ejection from the birth cluster shortly after formation. 
However, the result is a discrepancy between the position of the star relative to evolution tracks in the sHRD (determined 
by solution \#4) on the one hand and the HRD (determined by the Gaia parallax, \#4a) on the other ($\sim$16 to 
17\,$M_\odot$). Moreover, the calculation of the mass from Eq.~3 of Paper~I and adopting the Gaia 
distance, yields a third mass value of $\sim$11\,$M_\odot$ (though with large error margins). The latter may require 
non-standard evolution. Overall, no consistent picture is obtained.
 
A hint for a solution may come from the high Gaia DR3 RUWE value of 2.457 for $\rho$~Leo, implying the parallax not to 
be reliable and pointing towards a possible binary nature. The star lies close to the ecliptic, such that occultations 
by the moon occur, which can be exploited to verify the binary hypothesis. A first investigation by 
\citet{deVeGe76} found $\rho$~Leo to be a close double star with a projected separation of $\rho$\,=\,2.9$\pm$0.1mas 
and a magnitude difference of 0.04$\pm$0.09\,mag (at position angle 276\fdg5). On the other hand, \citet{EvEd81} found 
no duplicity and \citet{Radicketal82} reported $\rho$~Leo as a possible binary with $\rho$\,=\,10.3$\pm$0.8\,mas and a 
magnitude difference of 1.07$\pm$0.23\,mag in the $V$ band (at a position angle of 109\fdg1). $\rho$~Leo was resolved 
more recently by speckle interferometry using a H$\alpha$ filter, finding the companion at a position angle of 
98\fdg3$\pm$7\fdg7 at a separation $\rho$\,=\,46.1$\pm$1.8\,mas, with a brightness difference of 1.5\,mag 
\citep{Tokovininetal10}. The authors noted that a SB subsystem is suspected. \citet{Levatoetal88} concluded that the 
star is radial velocity variable, see their discussion for the long history of measurements (note that $\rho$~Leo shows 
non-radial oscillations), whereas \citet{Chinietal12} even characterised the star as SB2 on the basis of two spectra, but 
without giving further details.
A binary solution would explain the unusual pure absorption profile of H$\alpha$ observed for $\rho$~Leo, unlike the 
other B1\,Iab supergiants in Fig.~\ref{fig:spect_lum_showcase_new} which show P-Cygni profiles. Two lower-mass weak-
wind stars of very similar spectral type would be required to remain inconspicuous in the SED 
(Fig.~\ref{fig:sed_plot}), closely aligned in radial velocity and showing similar rotational velocities.
The presence of two continua would weaken the spectral lines, letting the Balmer lines appear narrower (thus yielding a Iab classification) as they in fact are and giving lower chemical abundances as are indeed present.

A further piece of the puzzle is the rotation period of 26.8\,d deduced from a 80\,d K2 lightcurve \citep{Aertsetal18}. 
From a comparison of $\varv_\mathrm{rot}$ (assuming a reasonable radius for the primary) and $\varv\sin i$ 
means that the system is seen not too far from being equator-on. The orbital motion is also likely to be co-
aligned, with the major axis roughly lying in the East-West direction based on the position angles cited above. In the 
absence of a reliable parallax, magnitude difference and separation of the components' lines\footnote{Several spectra 
of $\rho$~Leo are available in the CFHT and ESO archives from different instruments, but they do not show SB2 
character. The data of \citet{Chinietal12} showing two line systems would be pivotal in this 
context to analyse both components,
e.g.~using our methodology in analogy to \citet{Irrgangetal14} or \citet{Gonzalezetal17,Gonzalezetal19}.} 
no firm conclusions on the 
binary system can be drawn. But we may make some estimations guided by the available data. Assuming a distance of 
700\,pc and a magnitude difference of the components of 1.5\,mag, luminosities of $\log L/L_\odot$\,$\approx$\,4.85 
and $\log L/L_\odot$\,$\approx$\,4.25 would result, indicating a supergiant primary of about 16 to 17\,$M_\odot$ and a 
secondary of about 11\,$M_\odot$ close to the TAMS. Such a configuration would be able to reach the current position 
high above the galactic plane within the lifetime of the supergiant, assuming dynamical ejection early after formation. 
A highly eccentric orbit with the above orientation could explain the detection or absence (when the two stars are too 
close) of the second source at particular times, it would naturally explain a very low relative radial velocity 
difference of both stars and only at periastron would the two line systems perhaps be separated enough for identification. 
A highly eccentric orbit would also disfavour mass exchange between the two components.

With the rough sketch of the binary nature we can only stress that the stellar parameter and 
abundance solutions for the star as summarised in Tables~\ref{tab:stellar_parameters} and 
\ref{tab:abundances} provide a rough indication of the true values. In particular the abundances will be 
underestimated. Further monitoring of $\rho$~Leo is certainly needed to constrain the orbit and the properties of the 
binary components.\\[-8mm]
\paragraph{HD~114199 (ID\#5).} The reclassification of HD~114199 from supergiant to a bright giant Be star based on 
spectral morphology was already discussed in Sect.~\ref{section:observations}. It shows a pronounced CNO mixing 
signature (Fig.~\ref{fig:cno_logplot}) which is expected for a massive fast rotator -- it can actually be expected 
to rotate faster than the models with initial angular velocity of 
$\Omega_{\mathrm{rot}}$\,=\,0.568\,$\Omega_{\mathrm{crit}}$ discussed by \citet{Ekstroemetal12}, see 
\citet{Georgyetal13} for such models, which, however, terminate at masses slightly below that of HD~114199. A systematic underestimation of the star's initial angular velocity and correspondingly lower mass (see our discussion in Sect.~\ref{section:evol_status}) may also explain the discrepancy between compared distances. Even a small reduction of the star's mass (e.g. by 1\,$M_\odot$) brings our estimation well within the mutual uncertainty limits with the parallactic value. Its 
position in the (spectroscopic) HRD (Fig~\ref{fig:hrd}) likely corresponds to the star being on the blueward-evolving 
part of the track very close to the TAMS. Consequently, HD~114199 may still be in its core H-burning phase, unlike 
the other stars of our sample.\\[-8mm]
\begin{table*}
\caption{Literature values for atmospheric parameters and elemental abundances of Sher~25.}
\label{table:parameters_sher25}
\centering  
{\small
\setlength{\tabcolsep}{.8mm}
\begin{tabular}{llllllccccclll}
\hline\hline
ID & $T_\mathrm{eff}$ & $\log g$      & $y$             & $\xi$        & &  \multicolumn{5}{c}{$\log$\,($X$/H)\,+\,12}                                    & & code              & source\\ \cline{7-11}
\# & kK               & (cgs)         & (by number)     & km/s & & C             & N             & O             & Mg            & Si\\
\hline
A & 22.3$\pm$1.0     & 2.60$\pm$0.10 & 0.10$\pm$0.02   & 15           & & 7.01$\pm$0.06 & 8.42$\pm$0.12 & 8.87$\pm$0.07 & 7.46          & 7.42$\pm$0.07 & & {\sc Tlusty}/H+He & \citet{Smarttetal02}\\
B & 21.5$\pm$1.0     & 2.60$\pm$0.20 & ...             & 20           & & 7.82$\pm$0.16 & 8.52$\pm$0.09 & 8.50$\pm$0.03 & 7.59$\pm$0.36 & 7.40$\pm$0.20 & & {\sc Tlusty}      & \citet{Hendryetal08}\\
C & 22.0$\pm$1.0     & 2.60$\pm$0.15 & ...             & 23           & & ...           & 8.74$\pm$0.15 & 8.51$\pm$0.05 & ...           & 7.61$\pm$0.08 & & {\sc Fastwind}    & \citet{Hendryetal08}\\
D & 21.0$\pm$1.0     & 2.50$\pm$0.20 & ...             & 15           & & 7.8           & 8.5           & 8.1           & ...           & ...           & & {\sc Cmfgen}      & \citet{Hendryetal08}\\
1 & 20.9$\pm$0.5     & 2.61$\pm$0.06 & 0.117$\pm$0.015 & 16$\pm$2     & & 8.01$\pm$0.05              &  8.73$\pm$0.02             &  8.64$\pm$0.02             & 7.42               & 7.81$\pm$0.04              & & {\sc ADS}         & this work\\
\hline
\end{tabular}
\tablefoot{
Abundance uncertainties for the metals are standard errors of the mean, as provided in the literature.
}}
\end{table*}
\paragraph{HD~152235 (ID\#6).} The star is a member of the massive star cluster \object{NGC 6231} in the Sco OB1 
association, its age being commensurate with the cluster age of 4-7\,Myr \citep[, derived using Ekstr\"om et al. tracks 
for rotating stars]{Sungetal13}. It was identified as an X-ray source \citep{Kuhnetal17}. This supergiant shows 
stellar parameters close to those of Sher~25 and HD~13854, but is remarkable for its low CNO mixing signature. One may suppose 
it to stem from a slowly-rotating main-sequence progenitor. However, if one scales its $\varv \sin i$ value to the 
star's ZAMS radius assuming angular momentum conservation, an initial rotational velocity of at least 
350\,km~s$^{-1}$ is deduced. A possible scenario to solve the conundrum may be that the supergiant is located in a 
binary system with large enough separation during the main-sequence evolution to avoid tidal interactions, so that 
the star may have indeed been a slow rotator initially. Main-sequence stars in detached eclipsing binaries are known 
to show CNO abundances close to pristine values \citep{Pavlovskietal23}. The expansion to 
giant and supergiant dimensions may then have allowed tides to become effective, spinning up the stellar envelope. Detection 
of a companion and detailed binary modelling would be necessary to prove the validity of such a scenario, which
is beyond the scope of the present paper. However, we note that an indication for a presence of a binary companion 
is given by a Gaia DR3 RUWE value of 1.357 for HD~152235. 

\begin{figure}
\centering

\includegraphics[width=.85\hsize]{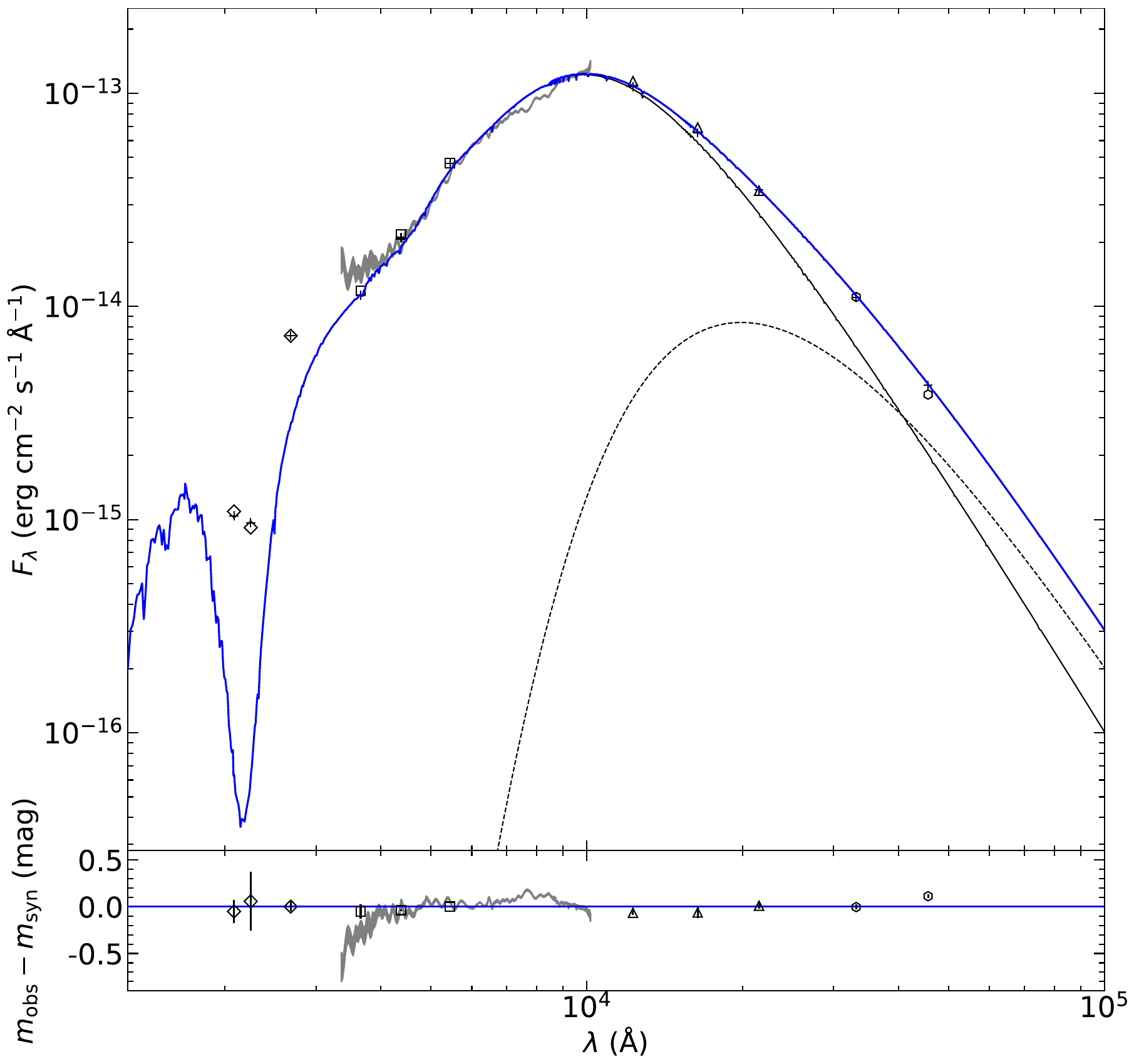}
      \caption{Spectral energy distribution of Sher~25. A comparison of the reddened model flux (solid blue line) to photometric and 
      spectrophotometric measurements (\textit{upper panel}) and the difference between modelled and observed 
      parameters measured in magnitudes (\textit{lower panel}) are shown. A blackbody contribution (black dashed line) with a temperature of $T$\,=\,1800\,K is assumed in addition to the stellar SED (black solid line). The grey line depicts Gaia spectrophotometry. Symbol assignment is the same as in 
      Fig.~\ref{fig:sed_plot}. In addition, near-UV Swift/UVOT photometric measurements (diamonds) are shown. 
      Results of synthetic photometry on the model SED in the respective pass bands are also indicated (plus signs).
      The error bars in the lower panel depict the $1\sigma$ uncertainty range.} 
         \label{fig:sed_sher25}
\end{figure}

\section{Discussion of Sher~25}\label{section:discussion}
The present analysis of Sher~25 finds similar atmospheric parameters to previous work that employed different non-LTE codes, as summarised in Table~\ref{table:parameters_sher25}. On the other hand, abundance values for the five elements studied so far showed a large scatter and seem strongly model dependent. Note that deviating from our usual notation, abundance uncertainties in Table~\ref{table:parameters_sher25} are standard errors of the mean, as provided in the literature. Our present values for CNO abundances show a high nitrogen enrichment and a depletion of both carbon and oxygen, relative to CAS values that should be representative for the initial CNO abundances because Sher~25 is located at a similar galactocentric radius as the solar neighbourhood. However, in contrast to some previous analyses -- which found a similar pattern -- our values together with solution~B align closely to the predicted mixing path in the N/O--N/C diagram (Fig.~\ref{fig:cno_logplot}). Solution~A from Table~\ref{table:parameters_sher25} provides a locus beyond the analytical CN-process boundary, while solution~D is inclined towards ON-processing. Solution~C is not discussed in this context, as no carbon abundance was provided. Concerning the heavier species, we recall that our silicon abundance is possibly overestimated (see Sect.~\ref{section:abundances_and_metallicity}).

\begin{figure*}[ht]
\sidecaption
    \includegraphics[width=12cm]{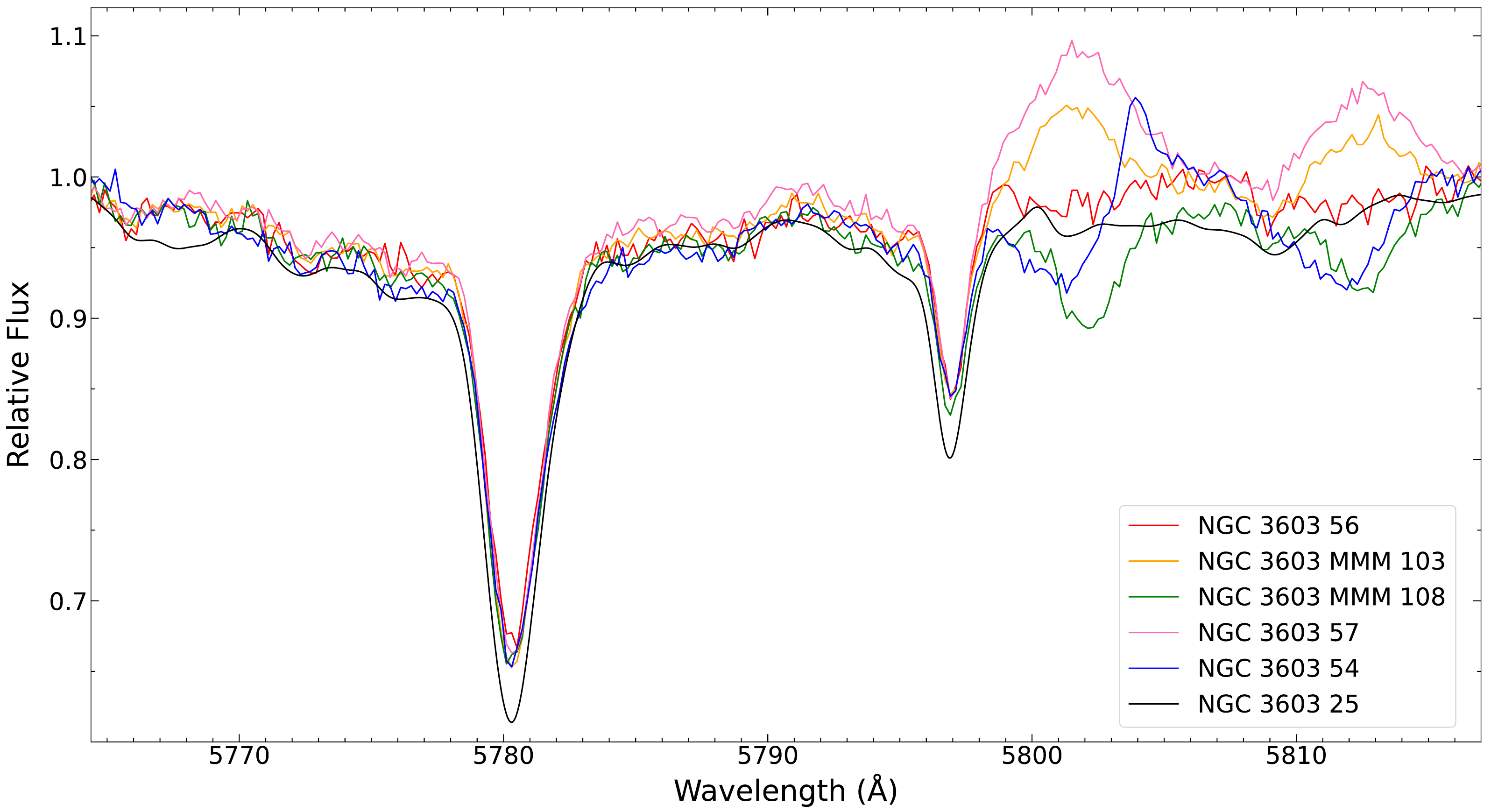}
    \caption{Spectra of Sher~25 and of five NGC~3603 bona-fide O-type cluster stars in the region of the two deep and narrow diffuse interstellar bands (DIBs) $\lambda$5780 and 5797\,{\AA}, see the legend. Other DIBs in this wavelength range are the shallow but broad $\lambda$5778\,{\AA} feature and the narrow and shallow $\lambda$$\lambda$5766, 5772, 5776, 5793, 5795 and 5809\,{\AA} features. The variation of the different spectra in the red is due to the different appearance of the stellar \ion{C}{iv} $\lambda$5801 and 5812\,{\AA} lines in the O-type stars. The spectrum of Sher~25 was artificially downgraded in resolution to match the lower $R$\,=\,5800 of the FLAMES/GIRAFFE data.}  
         \label{fig:dibs_comparison}
\end{figure*}

An analysis of Sher~25 and its hourglass nebula requires an understanding not just of the star itself, but also of its environment. Since this means understanding its relationship with the massive cluster \object{NGC 3603}, we have to accurately characterise the complex sight line towards Sher~25 to precisely define its spectroscopic distance. Again, fits to the the observed SED are employed, which is shown in Fig.~\ref{fig:sed_sher25}. Valuable UV photometry from the UVOT instrument on board the Swift mission are available (the available measurements in the UVW1, UVM2 and UVW2 bands were averaged) in addition to optical and IR photometry, and the spectrophotometry from Gaia. A comparison between observed and synthetic magnitudes is explicitly made. The synthetic photometry is based on filter curves and effective wavelengths adopted from the SVO Filter Profile Service\footnote{\url{http://svo2.cab.inta-csic.es/theory/fps/}} \citep{svoI,svoII}. The lower panel of Fig.~\ref{fig:sed_sher25} shows the residuals.
 
Good agreement between the model and observations is found for wavelengths below $\sim$10$^4$\,{\AA} for $E(B-V)$\,=\,1.66$\pm$0.03 and $R_V$\,=\,3.4$\pm$0.1\footnote{For comparison, \citet{Smarttetal02} adopted $E(B-V)$\,=\,1.6 and $R_V$\,=\,3.7$\pm$0.5, and 6.3$\pm$0.6\,kpc for the distance. \citet{Hendryetal08} employed the data of \citet{Melenaetal08}, who found a two-component reddening law with $E(B-V)$\,=\,1.1 and $R_V$\,=\,3.1 for the foreground and $E(B-V)$\,=1.39 and $R_V$\,=\,4.3 within the cluster, adding up to a total extinction of $A_V$\,=\,$3.1\times1.1 + [(E(B -V)-1.1)\times 4.3]$\,=\,4.657, and a distance of 7.6\,kpc. An overview of previous reddening and distance data toward NGC~3603 can be found in Table~4 of \citet{Melenaetal08} and in Table~A.1 of \citet{MaizApellanizetal20}.}, except for the Gaia spectrophotometry towards the blue and red limits (as also found for the comparison stars). On the other hand, the stellar SED (solid black line) is clearly insufficient to explain the marked observational IR excess (present in the $J$, $H$, and $K$ filters, as well as the WISE bands), which is also indicated by the IR glow in the area visualised in Fig.~\ref {fig:ngc3603_field}. This discrepancy can be removed by introducing a blackbody (BB) emitter with a temperature of $T$\,=\,1800\,K (dashed line in Fig.~\ref{fig:sed_sher25}), the nature of which cannot be constrained here. As such a BB emitter is necessary in order to reproduce the IR excess of some O-type stars in NGC~3603 as well (see Appendix~\ref{appendix:A}), it appears to be connected to NGC~3603, and not a circumstellar feature.
It may be due to hot dust, at the limiting temperature for sublimation of graphite grains, the most stable dust species.

This constrains our spectroscopic distance of Sher~25 to $d_{\mathrm{spec}}$\,=\,5440$\pm$700\,pc, whereas the Gaia-based distance is $d_{\mathrm{Gaia}}$\,= 5740$^{+800}_{-420}$\,pc, with a RUWE value of 0.943 indicating a reliable value. As discussed in detail in Appendix~\ref{appendix:B}, this location could put Sher~25 in principle in the vicinity of NGC~3603, which was calculated to to lie at a distance of $d_\mathrm{NGC3603}$\,=\,6250$\pm$140\,pc. However, the sheer existence of the hourglass nebula requires Sher~25 to be distant enough in the radial direction for this fragile structure not to be exposed to the strong winds of the WR- and early O-star population of NGC~3603. These stars have cleared the surrounding ISM in the lateral direction to larger distances than Sher~25 is located from the cluster core in projection and they have sculpted the pillar out of the molecular cloud in the south-eastern direction (see Fig.~\ref{fig:ngc3603_field}) at even larger lateral distance. A foreground location of Sher~25, as suggested by our spectroscopic solution, is therefore more likely.

A second approach towards a clear understanding of the relationship between the cluster and Sher~25 may be attempted by looking at information contained in lines of interstellar absorption. Figure~\ref{fig:dibs_comparison} shows the wavelength region around 5790\,{\AA}, which includes two of the strongest narrow diffuse interstellar bands (DIBs). In fact, practically the entire wavelength region, which for a normal B1 supergiant would be void of stellar lines (except for \ion{N}{ii} $\lambda$5767.4\,{\AA} in N-rich objects, such as in Sher~25), is dominated by DIB absorption because of the high extinction along the sight line. This includes in particular the shallow but very broad DIB $\lambda$5778\,{\AA} and a number of other features ($\lambda$$\lambda$5766, 5772, 5776, 5793, 5795 and 5809\,{\AA}) which are shallow and narrow. The plot also shows spectra of five O-type stars, assumed to be members of NGC~3603, taken with the GIRAFFE spectrograph of the VLT/FLAMES facility \citep{Pasquinietal02}. Note that the FEROS spectrum of Sher~25 was artificially downgraded here to match the $R$\,=\,5800 of the GIRAFFE data. In particular, the two strongest DIBs at $\lambda\lambda$5780 and 5797\,{\AA} distinctly show the disparity in strengths between Sher~25 on the one hand and the cluster members on the other: stars assumed to belong to NGC3603 show a consistently lower amount of DIB absorption. This is in spite of the variation in reddening \citep[the values fall in the range $E(B-V)$\,=\,1.30--1.42, see][]{Melenaetal08} and spread in angular distance $\theta$ from the cluster centre (i.e. $\theta$\,$\approx$\,8\arcsec~east for NGC~3603~MMM~108 to $\theta$\,$\approx$\,25\arcsec~west for NGC~3603~54). While this finding maintains our conclusion that the star is clearly not associated with the cluster, on a first glance it may contradict the localisation in the foreground. One may argue that Sher~25 should be located in the background of the cluster. However, what is required is a higher column density of DIB carriers to produce the stronger DIBs, which in addition to a longer path length may also be achieved by having a surplus of DIB carriers located in the circumstellar medium around Sher~25 (which would have interesting implications). Further investigations are certainly required, but are beyond the scope of the present work. In view of the previous discussion we therefore maintain our conclusion that Sher~25 is not associated with NGC~3603 and likely stands in the foreground of the cluster.

A distinct spatial separation of Sher~25 from NGC~3603 is also indicated by the spectra of the circumstellar hourglass nebula of Sher~25 and of the \ion{H}{ii} region surrounding NGC~3603. We refer to the analysis of the two nebular sites presented by \citet{Hendryetal08}, using the 'direct' method \citep[e.g.][]{Skillman98}. The cluster NGC~3603 is one of the most massive very young ($\sim$1--2\,Myr) star clusters known in the Milky Way. It is dominated by the light of early O-type stars and four WN6h stars, which provide a harsh UV radiation field that leads to high excitation in the surrounding giant \ion{H}{ii} region. This gives rise to strong [\ion{O}{iii}] $\lambda\lambda$4959 and 5007\,{\AA} emission, with the intensity of the weaker component showing about twice the flux of H$\beta$, see Figs.~4 and 5 of \citet{Hendryetal08}. This high excitation is much weaker in the spectra of the hourglass nebula, with [\ion{O}{iii}] showing only 0.6$\times$ the flux of H$\beta$, despite Sher~25 being closer in lateral projection to the cluster core than the \ion{H}{ii} clouds investigated by \citet{Hendryetal08} and despite the regions having about the same oxygen abundance. Instead, the spectrum of the hourglass nebula shows stronger lines from lower-excitation species. In consequence, Sher~25 is not located physically close to NGC~3603, but must be sufficiently separated in the radial direction to avoid exposure to the cluster's UV radiation field (i.e.~at a position in front of the cluster as deduced above). Adopting our spectroscopic distance and extinction value, the luminosity of Sher~25 becomes $\log L/L_\odot$\,=\,5.48$\pm$0.13, which for a single star scenario \citep[assuming evolutionary tracks for rotating stars from][]{Ekstroemetal12} corresponds to a star with $\sim$27\,$M_\odot$ on the ZAMS. The age of Sher~25 is $\sim$7-8\,Myr, about four to eight times the age of NGC~3603.

A comparison of nebular abundances for the NGC~3603 giant \ion{H}{ii} region and the hourglass nebula, as well as of the photospheric abundances of Sher~25 is made in Table~\ref{table:nebular_abundances}. The distance of NGC~3603 to the galactic centre is similar to that of the solar neighbourhood. As a consequence one would expect the giant \ion{H}{ii} region abundances to be close to CAS values (Table~\ref{tab:abundances}). Indeed, the heavier species Ne, S and Ar show such agreement, whereas N and O appear underabundant by $\sim$0.2--0.3\,dex. The origin of the latter in particular is not clear, but we note that some assumptions also had to be made for the nebular abundance determination \citep[see][ for details]{Hendryetal08}, such that additional unaccounted systematic uncertainties may need to be considered. The abundances of Sher~25 and the hourglass nebula are consistent within the error bars, with stellar ratios for N/C\,$\approx$\,5.2 and N/O\,$\approx$\,1.2 (by number). Besides the high degree of CNO-cycling, the hourglass nebula appears to show increased abundances of neon (also present in Sher~25) and of argon. 

\begin{table}
\caption{Comparison of nebular and photospheric abundances.}
\label{table:nebular_abundances}
\centering  
{\small
\setlength{\tabcolsep}{.7mm}
\begin{tabular}{lccccc}
\hline\hline
        & \multicolumn{5}{c}{$\log$\,($X$/H)\,+\,12}\\
 \cline{2-6}  
 Object & N & O & Ne & S & Ar\\
\hline
\ion{H}{ii} region\tablefootmark{a} & 7.47$\pm$0.18 & 8.56$\pm$0.07 & 8.11$\pm$0.09 & 7.22$\pm$0.22 & 6.44$\pm$0.08\\
hourglass\tablefootmark{a}          & 8.91$\pm$0.15 & 8.61$\pm$0.13 & 8.46$\pm$0.21 & ...           & 6.70$\pm$0.17\\
Sher~25\tablefootmark{b}            & 8.73$\pm$0.09 & 8.64$\pm$0.08 & 8.31$\pm$0.08 & 7.47$\pm$0.03 & ...\\
\hline
\end{tabular}
\tablefoot{
\tablefoottext{a}{\citet{Hendryetal08}}
\tablefoottext{b}{this work}
}}
\end{table}

Finally, the origin of the hourglass nebula needs to be addressed. In terms of evolutionary stage, Sher~25 is unlikely to have gone through a previous RSG phase according to its CNO mixing signature, as convective dredge-up would have brought the star to the top of the theoretical mixing signature curve in the single-star evolutionary scenario in Fig.~\ref{fig:cno_logplot}, in analogy to the discussion of HD~14956 (ID\#3) in the previous section. As a result, Sher~25 has expelled the hourglass nebula during its BSG phase with a dynamical age of about 6600\,yr \citep{Brandner97b}. The mass loss of rotating stars is of necessity anisotropic, with a fast wind emerging from the hotter polar caps, while an ejection of an equatorial ring may occur if the opacity in these regions grows rapidly for decreasing effective temperature. The mass loss may reach large values if a star reaches the rotationally-modified Eddington limit, the so-called $\Omega\Gamma$-limit \citep{MaMe00}. The effect is assumed to play an important role in the ejection of LBV nebulae \citep{Lamersetal01}. After our revision of its mass, Sher~25 is clearly unrelated to LBVs: it is significantly less luminous than the classical LBVs and much hotter than the low-luminosity LBVs. Can the $\Omega\Gamma$-limit have played a role in the formation of the hourglass nebula? Combining our $\varv\sin i$ of 60\,km\,s$^{-1}$ with the inclination angle of 64\degr derived by \citet{Brandner97a} by assuming an intrinsic circular ring geometry (in the star's equatorial plane), one finds a rotational velocity of 67\,km\,s$^{-1}$ for Sher~25. This is about three times higher than the rotational velocity implied by an (interpolated) 27\,$M_\odot$ Geneva model with an initial $\Omega_\mathrm{rot}$\,=\,$0.568\Omega_\mathrm{crit}$ for the parameters of Sher~25. Near the TAMS the rotational velocity for the star, which was more compact at that time, would have been about twice as large, still very far away from critical rotation -- moreover, Sher~25 is far from the Eddington limit. Consequently, the star was never close to the $\Omega\Gamma$-limit (note that two of our sample stars show a similar $\varv\sin i$, so Sher~25 is not unusual in this regard).

Alternatively, bipolar outflows including hourglass nebulae are often found for planetary nebulae, considered to be formed during the common-envelope phase in a binary system \citep[e.g.][, and references therein]{Ondratscheketal22}. Such a scenario was also invoked to produce the triple ring nebula around the precursor of SN~1987A \citep{Podsiadlowski92,MoPo07}. In order to reproduce the dynamics of the hourglass nebula around Sher 25 and the total ejected mass of 0.3 to 0.6\,$M_\odot$ \citep{Brandner97b} and the offset of the equatorial ring from the central star \citep{Brandner97a}, \citet{MoPo09} suggested that a binary merger had occurred during the Hertzsprung crossing when the envelope had a radius $\sim$300\,$R_\odot$, that is before the RSG phase, as indicated by the CNO mixing signature. The occurrence of the merger and not orbital shrinkage was backed by \citet{Tayloretal14}, who found Sher~25 to be a single star. They attributed some small-scale radial velocity variations to pulsations. Sher~25 may still be in a thermal readjustment phase after the merger and nebula ejection some 6600\,yr ago. Consequently, only our atmospheric parameters, the luminosity and radius may be firm, whereas the mass and age -- which were derived by considering (single star) evolution models -- need to be treated with caution. However, it still may be safe to assume Sher~25 to be significantly older than NGC~3603, while its mass may be even lower than inferred in the single-star scenario.

We conclude that our work finds Sher~25 to be much closer in properties to Sk$-$69{\degr}202, the precursor of SN~1987A, than previously assumed. It is therefore also more similar to the other two B1 supergiants with bipolar nebulae, [SBW2007] 1 \citep{Smithetal07} and \object{MN18} \citep{Gvaramadzeetal15}, and the cooler analogues HD~168625 \citep[B6\,Iap,][]{Smith07} and \object{HD 93795} \citep[B9\,Ia,][]{Gvaramadzeetal20}, at similar luminosities.

\begin{acknowledgements}
D.W., A.E. and N.P. gratefully acknowledge support from the Austrian Science Fund FWF project DK-ALM, grant 
W1259-N27. We thank M.~Urbaneja and S.~Kimeswenger for valuable discussions. 
We are grateful to A.~Irrgang for several updates of {\sc Detail} and {\sc Surface}. We want to thank the referee for valuable suggestions to improve the paper.
Based on observations collected at the European Southern Observatory under ESO programmes 075.D-0103(A) -- PI Dufton, 082.D-0136(A) -- PI Evans, 091.D-0221(A) -- PI Przybilla, obtained from the ESO Science Archive Facility with DOI: \url{https://doi.org/10.18727/archive/24}, and on programme 381.D-0914(A) -- PI Rochau, with DOI: \url{https://doi.org/10.18727/archive/27}. Based on observations collected at the Centro Astron\'omico Hispano 
Alem\'an at Calar Alto (CAHA), operated jointly by the Max-Planck Institut f\"ur  Astronomie and the Instituto de 
Astrof\'isica de Andaluc\'ia (CSIC), proposal H2005-2.2-016.
The latter observational data are available under \url{https://doi.org/10.5281/zenodo.8230158}.
This research used the facilities of the Canadian Astronomy Data Centre operated
by the National Research Council of Canada with the support of the Canadian Space Agency.
This work has made use of data from the European Space Agency (ESA) mission
{\it Gaia} (\url{https://www.cosmos.esa.int/gaia}), processed by the {\it Gaia}
Data Processing and Analysis Consortium (DPAC,
\url{https://www.cosmos.esa.int/web/gaia/dpac/consortium}). Funding for the DPAC
has been provided by national institutions, in particular the institutions
participating in the {\it Gaia} Multilateral Agreement.
This publication makes use of data products from the Two Micron All Sky Survey, which is a joint project of the University of Massachusetts
and the Infrared Processing and Analysis Center/California Institute of Technology, funded by the National Aeronautics and Space 
Administration and the National Science Foundation.
This publication makes use of data products from the Wide-field Infrared Survey Explorer, which is a joint project of 
the University of California, Los Angeles, and the Jet Propulsion Laboratory/California Institute of Technology, funded 
by the National Aeronautics and Space Administration. This research has made use of the SVO Filter Profile Service  (\url{http://svo2.cab.inta-csic.es/theory/fps/}) supported from the Spanish MINECO through grant AYA2017-84089.
\end{acknowledgements}

\clearpage

%
%

\typeout{}
\bibliographystyle{aa}
\bibliography{biblio.bib}

\begin{appendix} 

\section{IR excess in NGC~3603}\label{appendix:A}
\begin{figure}
\centering
\includegraphics[width=.9\hsize]{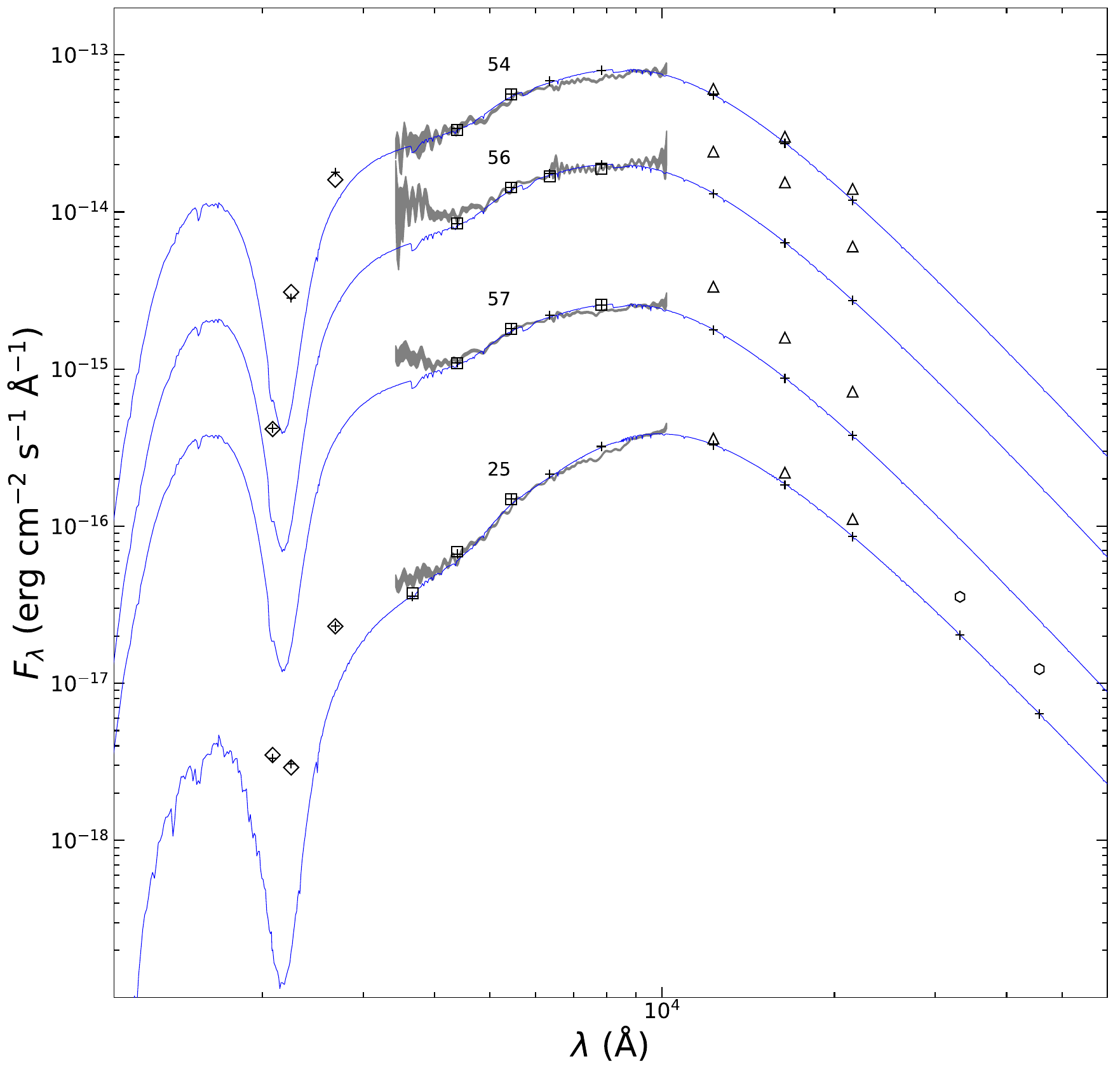}
      \caption{Comparison of stellar model SEDs and photometry for a subset of the cluster members depicted in Fig.~\ref{fig:dibs_comparison}. Symbol assignment is analogous to Fig.~\ref{fig:sed_sher25}. Johnson-Cousins $R_{\mathrm{c}}$ \& $I_{\mathrm{c}}$ are shown as black squares. For better visibility, the SEDs were shifted by $+1$\,dex for NGC~3603~54, $-1$\,dex for NGC~3603~57, and $-2.5$\,dex for Sher~25.} 
         \label{fig:cluster_seds}
\end{figure}

Sher~25 shows a notable observational IR excess, present already in the 2MASS $J$, $H$, and $K$ filters, as well as the WISE bands. While some excess flux in W2 is also present in most comparison stars, this behaviour is unusual for $J$, $H$, $K$, and W1 (see the discussion in Sect.~\ref{section:reddening_law}). As can be seen in Fig.~\ref{fig:cluster_seds}, a similar excess is also observed for some bona-fide cluster members of NGC~3603 (see discussion in Sect.~\ref{section:discussion}). For a useful comparison, model stellar spectra -- corresponding to spectral type O3\,V for NGC~3603~56 and 57, and O6\,V for NGC~3603~54 -- were downloaded from the Castelli-Kurucz atlas\footnote{\url{https://www.stsci.edu/hst/instrumentation/reference-data-for-calibration-and-tools/astronomical-catalogs/castelli-and-kurucz-atlas}}. For NGC~3603~56 and 57, which are located in a region with stronger background glow, it is apparent that an accurate fitting of the UV and optical photometry precludes a reproduction of the IR data without assumption of an extra component. A very mild excess is present in NGC~3603~54, likely due to its location in a region with almost absent background glow.

\section{The Gaia DR3 distance to NGC~3603}\label{appendix:B}
The sight line towards NGC~3603 is complicated for distance determinations because it traverses the Carina spiral arm 
-- one of the most active regions of massive star formation in the Milky Way -- close to tangentially over a 
large region. Chance projections of foreground or background early-type stars on the cluster area can therefore be 
expected. 

A comprehensive overview of distances to NGC~3603 from the literature that predated parallax measurements by the Gaia 
mission is given in Table~A.1 of \citet{MaizApellanizetal20}. Most values from the 18 studies lie in the range 6.0 to 
8.5\,kpc, with some outliers. \citet{MaizApellanizetal20} themselves derived a Gaia DR2-based distance of 
8000$^{+2600}_{-1700}$\,pc, which was recently updated to a value of 
7130$^{+590}_{-500}$\,pc 
\citep{MaizApellanizetal22} based on Gaia EDR3 data. A total of 166 stars contributed to the latter value
after filtering an initial number of about 28\,600 stars in the area of NGC~3603 within a circle centred on 
right ascension $\alpha$\,=\,168{\fdg}79 and declination $\delta$\,=\,$-$61{\fdg}26 with a radius of 206{\arcsec}
according to the Renormalised Unit Weight Error (RUWE), position, proper motion and colour. 

\begin{figure*}
\sidecaption 
    \resizebox{12cm}{!}{\includegraphics{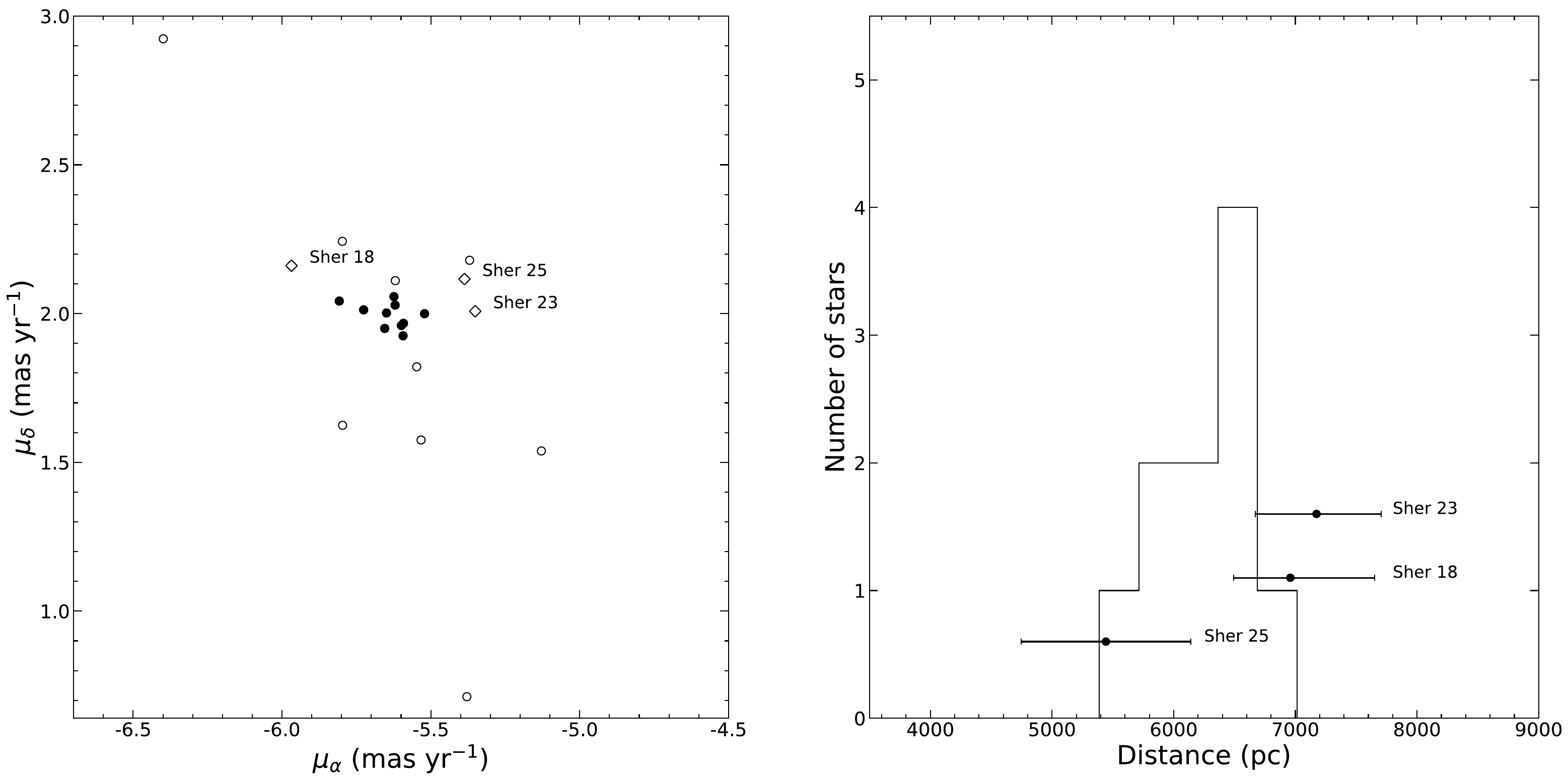}}
      \caption{Gaia EDR3 proper motions \citep{Gaia2020} and photogeometric distances \citep{Bailer-Jones_etal_2021} of presumed member stars of NGC~3603 identified in Table~\ref{table:ngc3603}. \textit{Left panel:} proper motions in right ascension $\mu_{\alpha}$ and in declination $\mu_{\delta}$. Objects depicted as full symbols satisfy our selection criteria, while those marked by empty symbols are excluded for high RUWE values $>$1.3 or constraints in proper motion. \textit{Right panel:} distribution of inferred distances towards NGC~3603. The BSGs in the inner NGC~3603 field (Sher~25, Sher~23, and Sher~18) are marked in both panels.} 
         \label{fig:ngc3603_histo}
\end{figure*}

As the uncertainties in the cluster distance are substantial -- note that active massive star formation in the wider 
area of NGC~3603 is ongoing \citep{RomanLopesetal16} with a possible substantial extension also in distance -- we 
tried a modified approach to improve on the cluster distance. We concentrated on the inner cluster 
(1\arcmin$\times$1\arcmin~field) as investigated by \citet{Melenaetal08}, employing only stars with Gaia EDR3 
parallaxes \textit{and} known (early) spectral type \citep[from Table~3 of][]{Melenaetal08}. The resulting star 
sample is summarised in Table~\ref{table:ngc3603}, which contains the star name, as resolved by 
SIMBAD\footnote{\url{http://simbad.u-strasbg.fr/simbad/sim-fid}}, the spectral type, Gaia EDR3 parallax $\varpi$ and 
proper motion components in right ascension $\mu_\alpha$ and declination $\mu_\delta$, the Gaia $G$ magnitude, the
RUWE and the photogeometric distance \citep{Bailer-Jones_etal_2021}. The latter values and the parallaxes show that 
Gaia measurements considering solely the first 34 months of the mission are strongly limited in accuracy and 
precision for such distant stars. From this star sample, further objects were 
removed for RUWE values $>$1.3, which is an indication of possible binarity, or because of discrepant proper 
motions, as visualised in Fig.~\ref{fig:ngc3603_histo}. The remaining stars are marked in boldface style in 
Table~\ref{table:ngc3603} for which a histogram of the distance distribution is also shown in 
Fig.~\ref{fig:ngc3603_histo}. We note that the two late-O and early-B supergiants \object{NGC 3603 23} and 25, 
aka Sher 23 (OC9.7\,Ia) and our sample star Sher 25 were also removed because they do not match the 1 to 2\,Myr 
isochrone of NGC~3603 \citep[Fig.~7 of ][]{Melenaetal08}, but instead appear to be significantly older. We also note that we do not expect crowding effects on the parallaxes, as all selected stars are isolated (cf.~Fig.~1 and 2 of Melena et al., and our Fig.~\ref{fig:ngc3603_field}) and separated by >\,8\arcsec~from the cluster centre.

The distance to NGC~3603 was consequently calculated from the remaining 10 stars to 
$d_\mathrm{NGC 3603}$\,=\,6250\,pc with a standard error of 150\,pc and a 1$\sigma$ standard deviation of 460\,pc. 
This is somewhat shorter than the 7130$^{+590}_{-500}$\,pc value of \citet{MaizApellanizetal22} but compatible 
within the mutual uncertainties.
A significant reduction of the uncertainties has to await further Gaia data releases based on a longer measurement
period and a better understanding of systematic effects.

Finally, we want to comment on the atmospheric parameters that were provided in the full Gaia DR3 \citep{GaiaDR3}.
Effective temperatures, surface gravities and metallicities are found for ten out of the 23 stars from 
Table~\ref{table:ngc3603}. All these confuse the O-stars and Sher~25 with metal-poor giants with 
$T_\mathrm{eff}$\,$<$\,10\,000\,K (with one exception at $\sim$17\,000\,K). Apparently, the Gaia astrophysical 
parameters inference system \citep[Apsis,][]{Fouesneauetal22} needs to be improved in order to
enable it to correctly characterise significantly reddened O-type stars, similar to the case of mildly-reddened late O-type stars \citep{Aschenbrenneretal23}.

\begin{table*}[!ht]
    \centering
    \caption{Gaia EDR3 parallaxes and photometric and astrometric measurements for stars in 
    NGC~3603.}\label{table:ngc3603}
    \small
    \renewcommand{\arraystretch}{1.2} 
    \begin{tabular}{l l r r r r r r}
    \hline
    \hline
    Name & Sp. Type\tablefootmark{a} & $\varpi$\,(mas) & $\mu_{\alpha}$\,(mas\,yr$^{-1}$) & $\mu_{\delta}$\,(mas\,yr$^{-1}$) & $G$\,(mag) & 
    RUWE & $d$\,(pc)\tablefootmark{b}\\
    \hline
NGC\,3603\,18         & O3.5\,If   & $0.1428 \pm 0.0111$ & $-5.968 \pm 0.012$ & $2.161 \pm 0.011$ & $11.9722$ & $0.99$  & $6960^{+690}_{-470}$ \\ 
NGC\,3603\,19         & O3\,V((f)) & $0.0897 \pm 0.0114$ & $-5.370 \pm 0.012$ & $2.179 \pm 0.012$ & $13.1764$ & $1.02$  & $7780^{+850}_{-630}$ \\ 
\textbf{NGC\,3603\,21}         & O6\,V((f)) & $0.1147 \pm 0.0160$ & $-5.593 \pm 0.019$ & $1.925 \pm 0.017$ & $14.2138$ & $1.00$  & $6400^{+790}_{-500}$ \\ 
\textbf{NGC\,3603\,22}         & O3\,III(f) & $0.1430 \pm 0.0170$ & $-5.808 \pm 0.018$ & $2.042 \pm 0.018$ & $12.7677$ & $1.07$  & $6470^{+770}_{-640}$ \\ 
NGC\,3603\,23                  & OC9.7\,Ia  & $0.1344 \pm 0.0112$ & $-5.351 \pm 0.012$ & $2.008 \pm 0.011$ & $12.1323$ & $0.98$  & $7170^{+530}_{-500}$ \\ 
\textbf{NGC\,3603\,24}         & O6\,V      & $0.1161 \pm 0.0152$ & $-5.592 \pm 0.016$ & $1.967 \pm 0.015$ & $13.7393$ & $1.13$  & $6130^{+570}_{-490}$ \\ 
NGC\,3603\,25                  & B1\,Iab    & $0.1560 \pm 0.0166$ & $-5.387 \pm 0.020$ & $2.116 \pm 0.015$ & $11.4261$ & $0.94$  & $5740^{+800}_{-420}$ \\ 	
\textbf{NGC\,3603\,27}         & O7.5\,V    & $0.1321 \pm 0.0182$ & $-5.599 \pm 0.022$ & $1.960 \pm 0.022$ & $14.5238$ & $1.03$  & $5390^{+600}_{-330}$ \\ 
NGC\,3603\,47         & O4\,V      & $0.1248 \pm 0.0103$ & $-5.797 \pm 0.011$ & $2.243 \pm 0.010$ & $12.1486$ & $0.94$  & $7910^{+580}_{-510}$ \\ 
NGC\,3603\,49                  & O7.5\,V    & $0.1845 \pm 0.0588$ & $-5.129 \pm 0.065$ & $1.538 \pm 0.060$ & $14.1456$ & $4.01$  & $3830^{+930}_{-430}$ \\ 
NGC\,3603\,53                  & O8.5\,V    & $0.2227 \pm 0.0466$ & $-5.619 \pm 0.053$ & $2.111 \pm 0.040$ & $13.9243$ & $3.25$  & $3330^{+830}_{-340}$ \\ 
\textbf{NGC\,3603\,54}         & O6\,V      & $0.1139 \pm 0.0147$ & $-5.655 \pm 0.016$ & $1.950 \pm 0.015$ & $14.0344$ & $0.97$  & $6360^{+490}_{-420}$ \\ 
\textbf{NGC\,3603\,56}         & O3\,III(f) & $0.1254 \pm 0.0216$ & $-5.620 \pm 0.022$ & $2.028 \pm 0.022$ & $12.9838$ & $1.04$  & $6660^{+1010}_{-460}$ \\ 
\textbf{NGC\,3603\,57}         & O3\,III(f) & $0.1521 \pm 0.0119$ & $-5.649 \pm 0.013$ & $2.002 \pm 0.013$ & $12.7444$ & $1.01$  & $6390^{+470}_{-430}$ \\ 
NGC\,3603\,63                  &O3.5\,III(f)& $0.0053 \pm 0.0407$ & $-5.533 \pm 0.055$ & $1.575 \pm 0.037$ & $12.9097$ & $1.91$  & $9300^{+1640}_{-1090}$ \\
\textbf{NGC\,3603\,64}         & O3\,V((f)) & $0.1097 \pm 0.0127$ & $-5.521 \pm 0.014$ & $2.000 \pm 0.012$ & $13.0839$ & $1.15$  & $7010^{+650}_{-480}$ \\ 
Cl* NGC 3603 MMM\,102          & O8.5\,V    & $0.1178 \pm 0.0602$ & $-5.379 \pm 0.066$ & $0.712 \pm 0.062$ & $14.8683$ & $2.57$  & $4450^{+710}_{-540}$ \\ 
\textbf{Cl* NGC 3603 MMM\,103} & O3\,V((f)) & $0.1583 \pm 0.0159$ & $-5.624 \pm 0.016$ & $2.057 \pm 0.027$ & $12.6349$ & $0.97$  & $5950^{+550}_{-440}$ \\ 
Cl* NGC 3603 MMM\,104          & O3\,III(f) & $0.0617 \pm 0.0442$ & $-5.548 \pm 0.065$ & $1.821 \pm 0.037$ &  $12.5400$ & $2.42$  & $7200^{+1110}_{-1010}$ \\ 
Cl* NGC 3603 MMM\,108          & O5.5\,V    & $-0.1445 \pm 0.1030$& $-5.797 \pm 0.111$ & $1.624 \pm 0.113$ & $13.2060$  & $5.83$  & $5200^{+2890}_{-1080}$ \\ 	
\textbf{Cl* NGC 3603 MMM\,117} & O6\,V      & $0.1043 \pm 0.0275$ & $-5.726 \pm 0.042$ & $2.012 \pm 0.025$ & $13.6948$ & $1.28$  & $5790^{+650}_{-560}$ \\ 
Cl* NGC 3603 BLW\,A2           & O3\,V      & $1.1375 \pm 0.4312$ & $-8.387 \pm 0.420$ & $1.162 \pm 0.419$ & $10.5626$ & $13.59$ & $750^{+1450}_{-110}$  \\ 
Cl* NGC 3603 BLW\,A3           & O3\,III(f*)& $0.5897 \pm 0.1579$ & $-6.399 \pm 0.174$ & $2.923 \pm 0.438$ & $10.7691$ & $4.22$  & $1420^{+300}_{-180}$ \\ 
\hline
    \end{tabular}
    \tablefoot{
    The objects used for the distance estimate of NGC~3603 are set in boldface style. We rejected the other 
    objects based on their RUWE factor or discrepant proper motions. The supergiants NGC 3603 23 and NGC~3603~25 (i. e.~Sher~25)
    were rejected because their ages are incompatible with the cluster isochrone of 1 to 2\,Myr 
    \citep{Melenaetal08}. Star identifications for the first 16 objects are from \citet{Sher65}, for the following five objects from \citet{Melenaetal08} and for the final two objects from \citet{Baieretal85}.
    \tablefoottext{a}{From Table~3 of \citet{Melenaetal08} except for NGC 3603 56, for which the spectral type of 
    the primary component from their Table~1~is~given.}
    \tablefoottext{b}{Photogeometric distances \citep{Bailer-Jones_etal_2021}.}
    }
\end{table*}

\section{Global model fit for Sher~25}\label{appendix:C}
A comparison of the observed FEROS spectrum of Sher~25 with the best fitting global synthetic spectrum is shown
in the following Figs.~\ref{fig:sher25_1} to \ref{fig:sher25_9}. The model was computed with the 
{\sc Atlas9/Detail/Surface} codes based on atmospheric parameters and elemental abundances as summarised in
Tables~\ref{tab:stellar_parameters} and \ref{tab:abundances}, respectively. With the exception of \ion{N}{iii} $\lambda$$\lambda$4634 and 4640\,{\AA}, all visible spectral lines of stellar origin are accounted for by the synthetic model spectrum. The diagnostic stellar lines are identified. We note that the lower Balmer lines, in particular H$\alpha$ and H$\beta$, but also H$\gamma$ to a lesser degree, as well as the strong red \ion{He}{i} $\lambda$5785, 6678 and 7065\,{\AA} lines are affected by the stellar wind, which is unaccounted for by the hydrostatic model. We also want to mention the difficulties in reproducing the \ion{C}{ii} $\lambda\lambda$4267 and 6578/82\,{\AA} lines closely, as addressed in Sect.~\ref{section:abundances_and_metallicity}. Multiple interstellar (’IS’) atomic lines, such as the \ion{Ca}{ii} H and K lines, the \ion{Ca}{i} $\lambda$4226\,{\AA}, the Na D lines, the \ion{K}{i} $\lambda$$\lambda$7664.9 and 7698.9\,{\AA} (the former partially overlaps with a telluric line of O$_2$) and also molecular absorption lines of CH $\lambda$4300 as well as CH$^+$ $\lambda$$\lambda$3957 and 4232\,{\AA}, are also identified. Note that they are blue-shifted in the observed spectrum because the stellar features were corrected to the laboratory rest frame. In addition, we emphasise the presence of numerous diffuse interstellar bands (DIBs) of considerable strength, mediated by the high reddening along the sight line towards Sher~25. These are absent in the model, as well as the numerous sharp telluric water vapour 
features and the A-, B- and $\gamma$-bands of O$_2$ towards the red part of the spectrum.

\begin{figure*}
\centering 
\includegraphics[angle=90,width=0.70\textwidth]{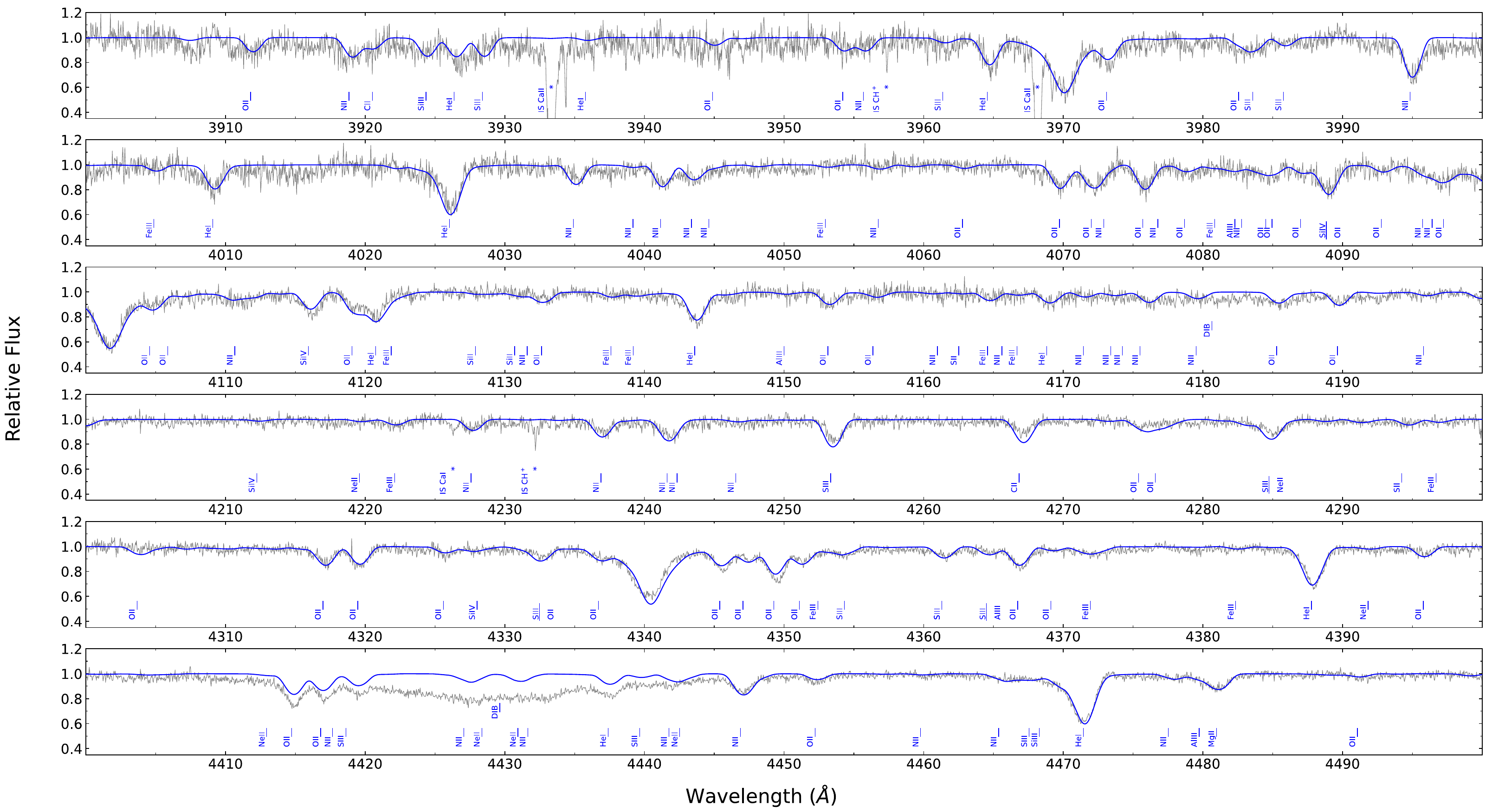}
        \caption{Comparison between the observed spectrum of Sher~25 (grey) and the best fitting synthetic spectrum (blue) in the wavelength range of 3900 to 4500\,{\AA}. The observed spectrum was shifted such that the stellar lines are in the laboratory rest frame.}
    \label{fig:sher25_1}
\end{figure*}

\begin{figure*}
\centering 
\includegraphics[angle=90,width=0.70\textwidth]{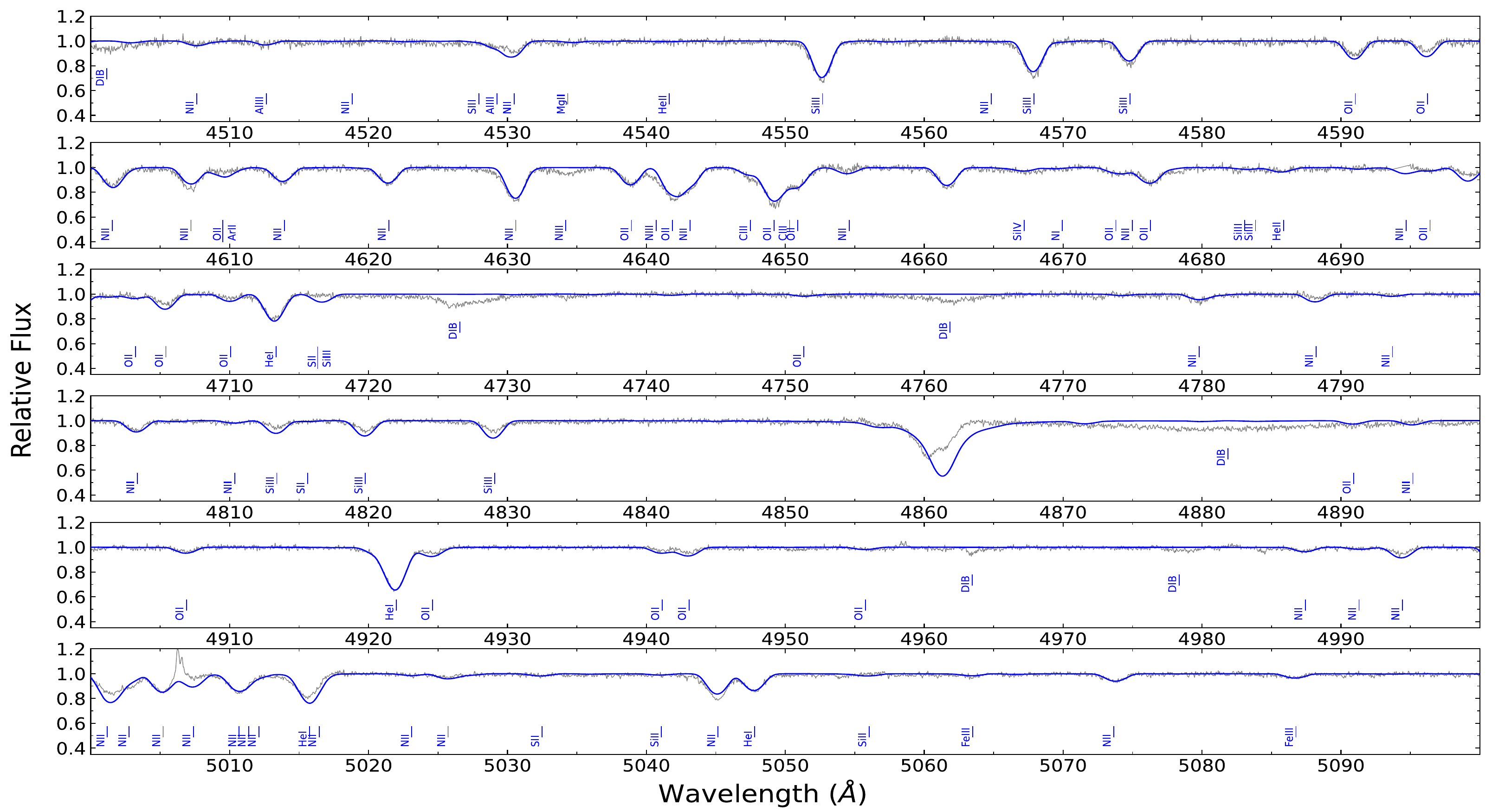}
        \caption{Same as Fig.~\ref{fig:sher25_1}, but in the wavelength range $\lambda\lambda$4500--5100\,{\AA}.}
    \label{fig:sher25_2}
\end{figure*}

\begin{figure*}
\centering 
\includegraphics[angle=90,width=0.80\textwidth]{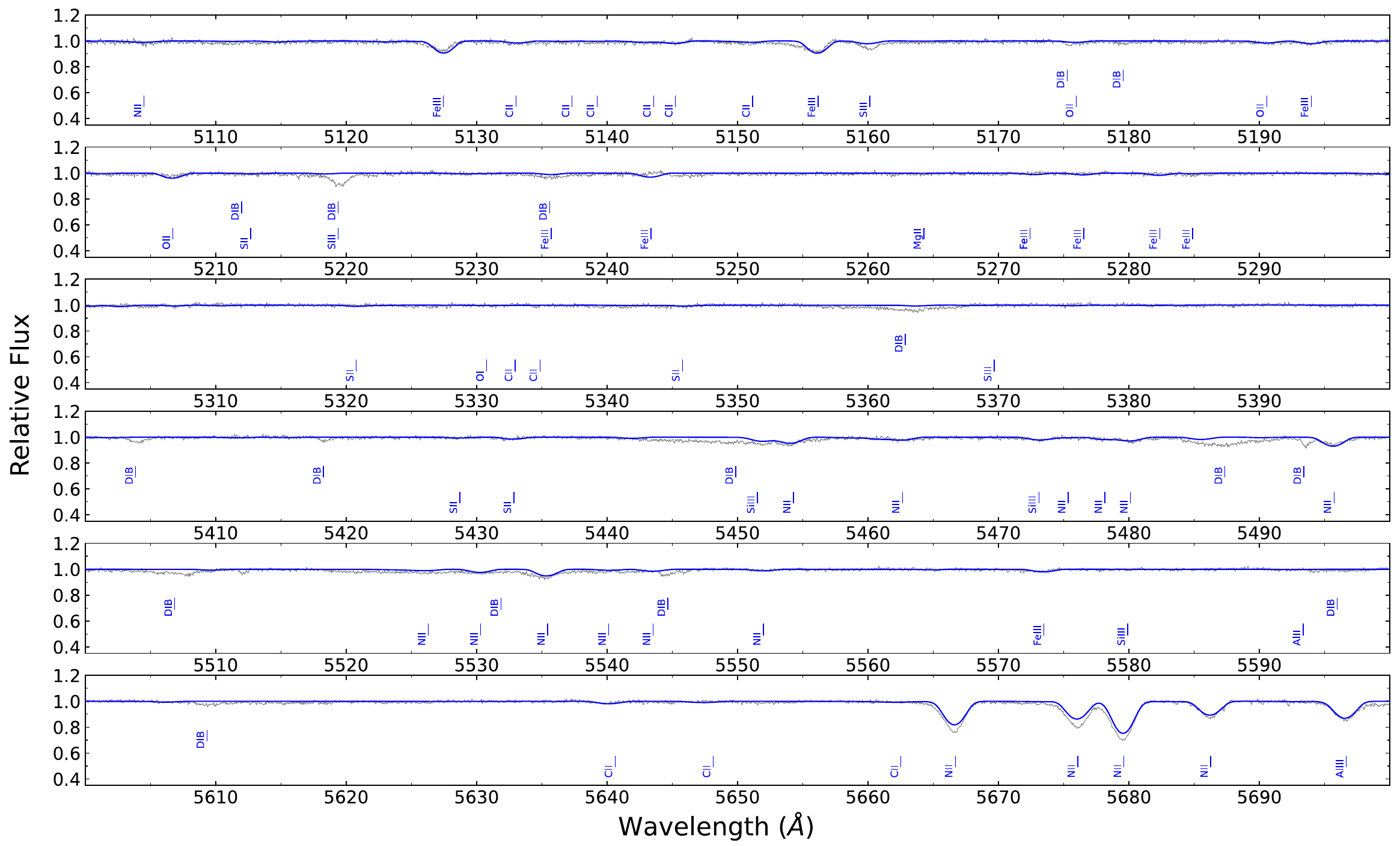}
        \caption{Same as Fig.~\ref{fig:sher25_1}, but in the wavelength range $\lambda\lambda$5100--5700\,{\AA}.}
    \label{fig:sher25_3}
\end{figure*}

\begin{figure*}
\centering 
\includegraphics[angle=90,width=0.80\textwidth]{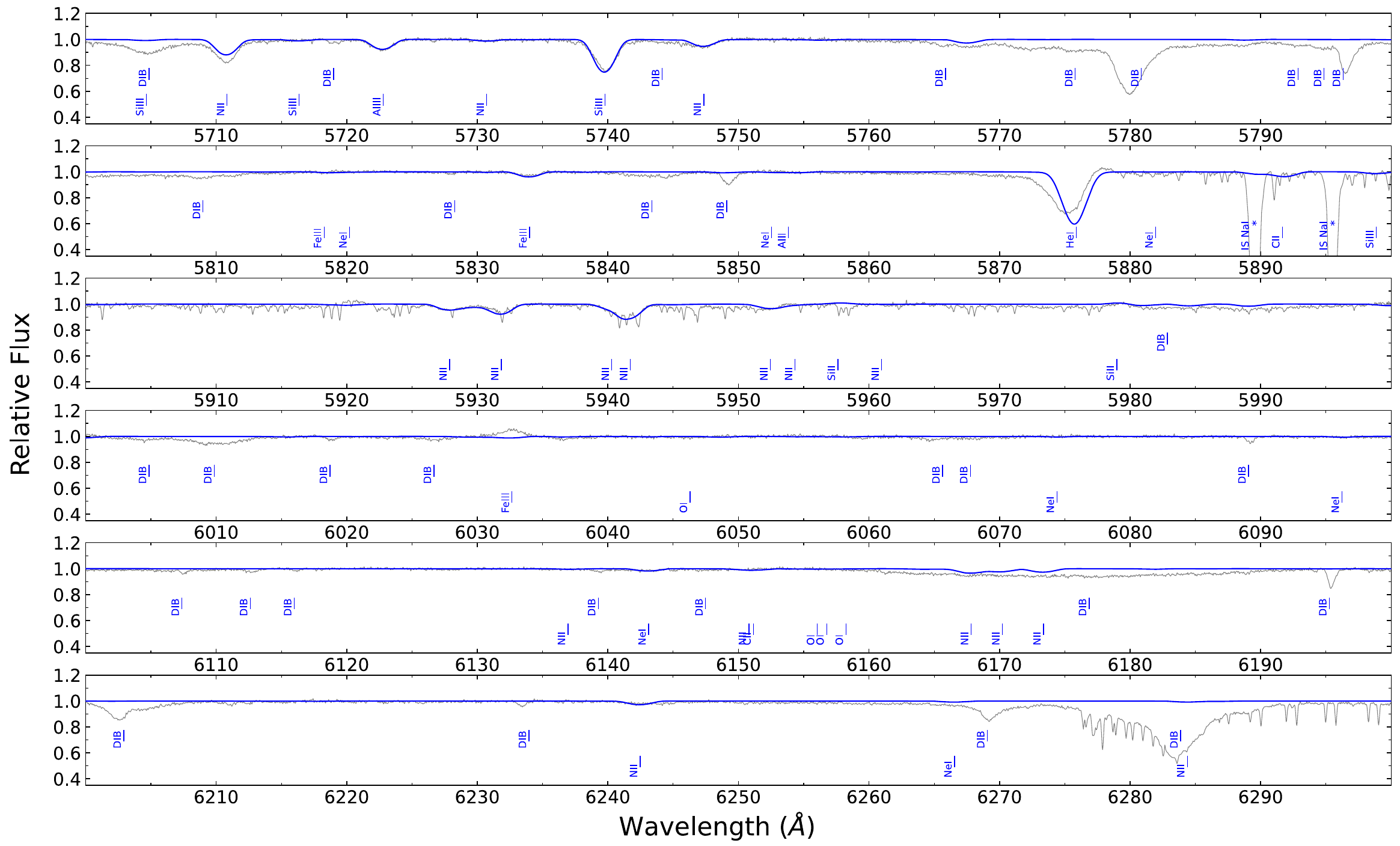}
        \caption{Same as Fig.~\ref{fig:sher25_1}, but in the wavelength range $\lambda\lambda$5700--6300\,{\AA}.}
    \label{fig:sher25_4}
\end{figure*}

\begin{figure*}
\centering 
\includegraphics[angle=90,width=0.80\textwidth]{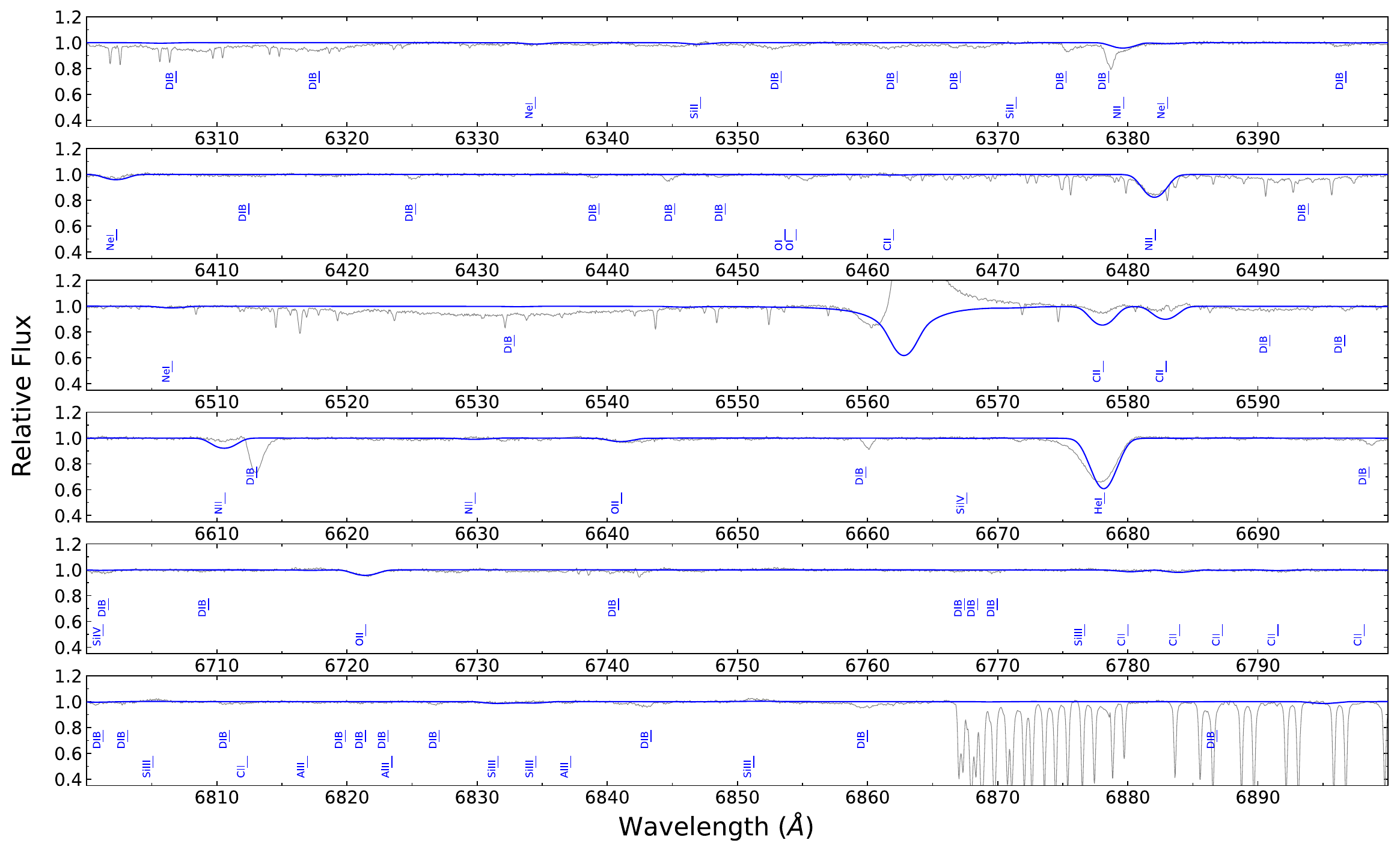}
        \caption{Same as Fig.~\ref{fig:sher25_1}, but in the wavelength range $\lambda\lambda$6300--6900\,{\AA}.}
    \label{fig:sher25_5}
\end{figure*}

\begin{figure*}
\centering 
\includegraphics[angle=90,width=0.80\textwidth]{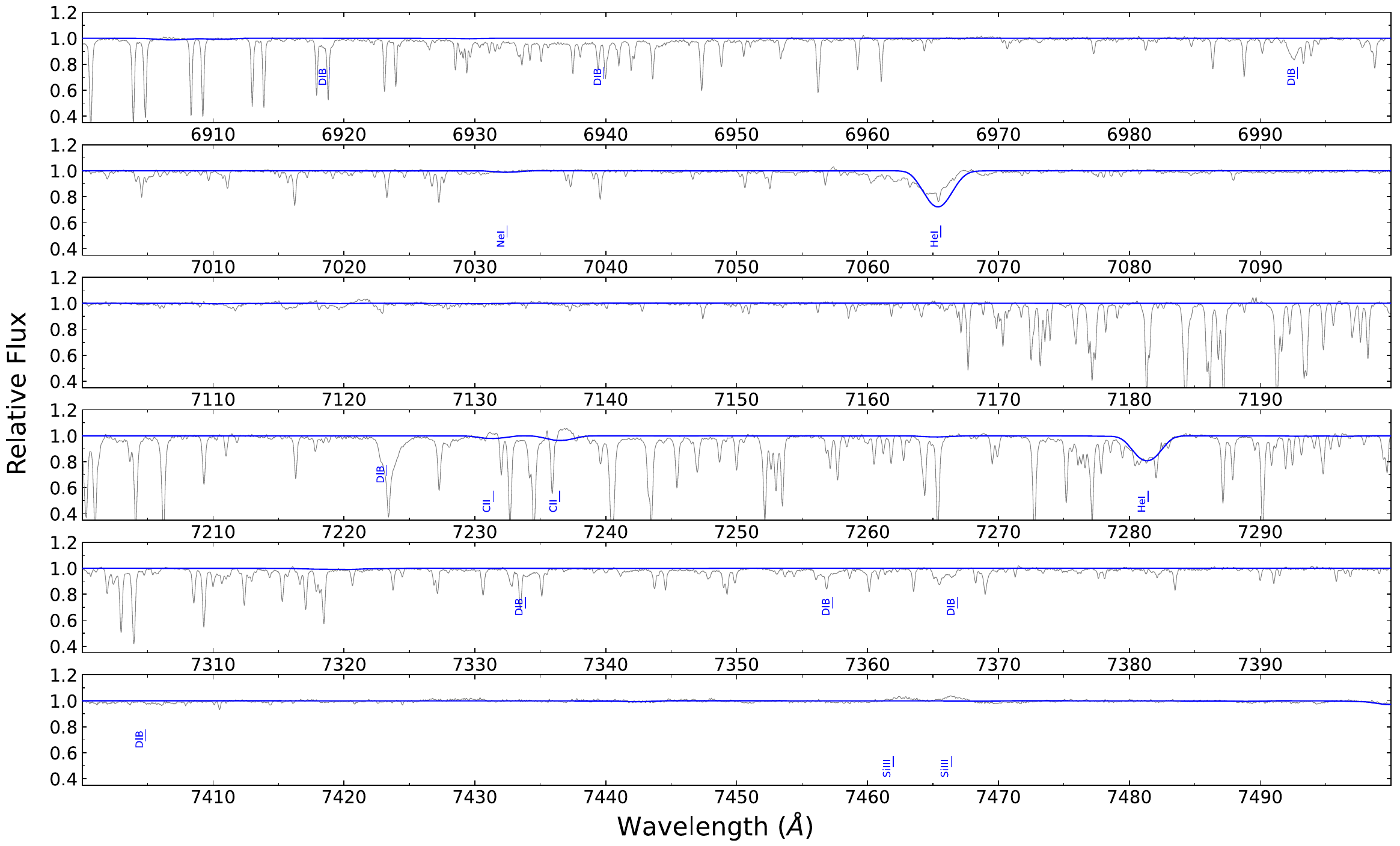}
        \caption{Same as Fig.~\ref{fig:sher25_1}, but in the wavelength range $\lambda\lambda$6900--7500\,{\AA}.}
    \label{fig:sher25_6}
\end{figure*}

\begin{figure*}
\centering 
\includegraphics[angle=90,width=0.80\textwidth]{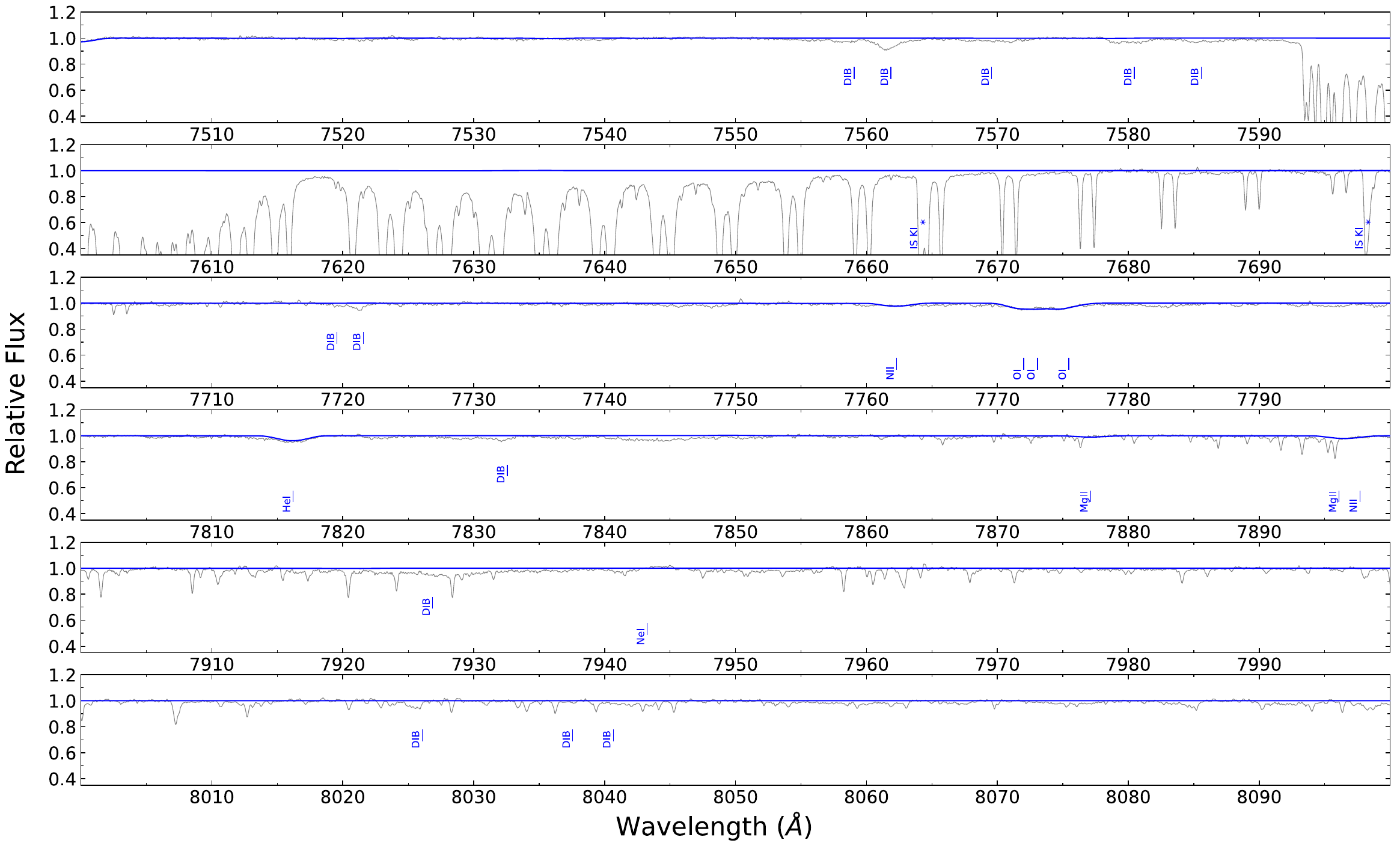}
        \caption{Same as Fig.~\ref{fig:sher25_1}, but in the wavelength range $\lambda\lambda$7500--8100\,{\AA}.}
    \label{fig:sher25_7}
\end{figure*}

\begin{figure*}
\centering 
\includegraphics[angle=90,width=0.80\textwidth]{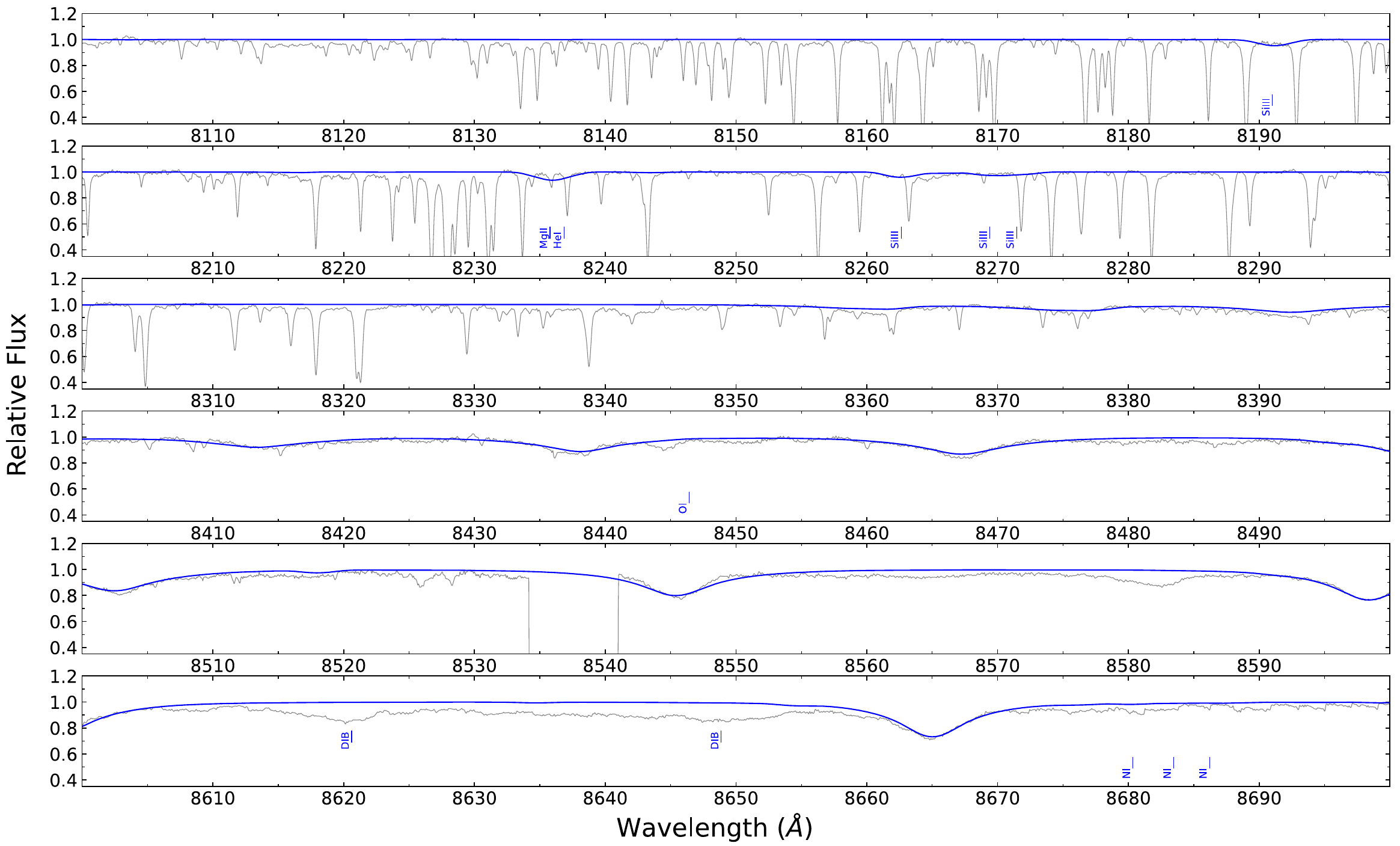}
        \caption{Same as Fig.~\ref{fig:sher25_1}, but in the wavelength range $\lambda\lambda$8100--8700\,{\AA}.}
    \label{fig:sher25_8}
\end{figure*}

\begin{figure*}
\centering 
\includegraphics[angle=90,width=0.80\textwidth]{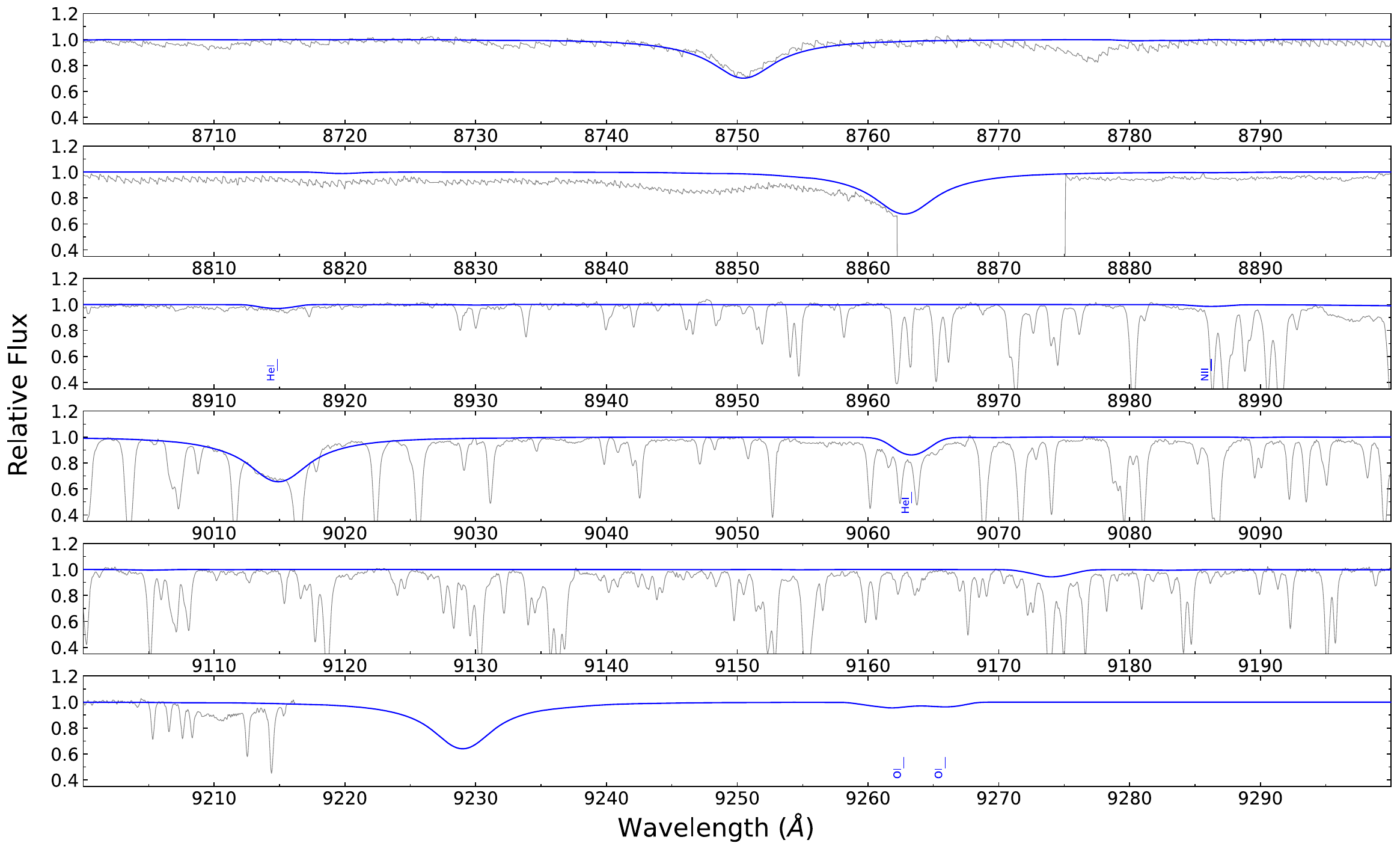}
        \caption{Same as Fig.~\ref{fig:sher25_1}, but in the wavelength range $\lambda\lambda$8700--9300\,{\AA}.}
    \label{fig:sher25_9}
\end{figure*}

\end{appendix}

\end{document}